\documentclass[letterpaper, 10 pt, conference]{IEEEtran}
  
\usepackage[utf8]{inputenc}
\usepackage{hyperref}
\PassOptionsToPackage{margin=0.73in}{geometry}
\usepackage[skip=0pt plus1pt, indent=0pt]{parskip}
\usepackage[english]{babel}
\usepackage{tikz}
\usetikzlibrary{positioning,shapes,shadows,arrows}
\usepackage[nottoc]{tocbibind}
\usepackage{amsfonts}
\usepackage{amsthm}
\usepackage{amsmath}

\usepackage{booktabs}
\usepackage{mathtools}
\usepackage{amssymb,epsfig,multirow}
\usepackage{amsbsy}
\usepackage[compact]{titlesec}
\titlespacing{\section}{0pt}{1ex}{1ex}
\titlespacing{\subsection}{0pt}{1ex}{0ex}
\titlespacing{\subsubsection}{0pt}{0ex}{0ex}
\usepackage{graphicx}
\usepackage{authblk}
\usepackage{geometry}
\usepackage{url}
\usepackage{float}
\usepackage{color}
\usepackage{enumerate}
\renewcommand{\baselinestretch}{1}
\usepackage[pdf]{graphviz}
\usepackage{tikz}
\usetikzlibrary{arrows.meta}
\usepackage{tkz-euclide}
\usepackage{tkz-graph}
\usetikzlibrary{graphs,quotes,arrows.meta}
\usepackage{subcaption}
\theoremstyle{plain}

\newtheorem{theorem}{Theorem}
\numberwithin{theorem}{section}
\newtheorem{lem}[theorem]{Lemma}
\newtheorem{cor}[theorem]{Corollary}

\newtheorem{prop}[theorem]{Proposition}

\theoremstyle{definition}
\newtheorem{defn}[theorem]{Definition}
\newtheorem{exam}[theorem]{Example}

\theoremstyle{remark}
\newtheorem{obser}[theorem]{Observation}

\newcommand{\G}{\mathcal{G}}
\newcommand{\A}{\mathcal{A}}
\newcommand{\T}{\mathcal{T}}
\newcommand{\C}{\mathcal{C}}

\allowdisplaybreaks

\title{\LARGE \bf Multi-Agent Path Finding on Strongly Connected Digraphs: feasibility and solution algorithms\vspace{-5pt}}

\author{S. Ardizzoni$^{1}$, I. Saccani$^{1}$, L. Consolini$^{1}$,  M. Locatelli$^{1}$  \\
       { \small $^{1}$Dipartimento di Ingegneria e Architettura, Universit\`a di Parma, Parco Area delle Scienze, 181/A, Parma, Italy}
}

\date{}
\begin{document}
	\tikzstyle{abstract}=[rectangle, draw=black, rounded corners, fill=blue!40, drop shadow,
	text centered, anchor=north, text=white, text width=3cm]
	\tikzstyle{comment}=[rectangle, draw=black, rounded corners, fill=red, drop shadow,
	text centered, anchor=north, text=white, text width=3cm]
	\tikzstyle{myarrow}=[->, >=open triangle 90, thick]
	\tikzstyle{line}=[-, thick]
	
	\maketitle
\begin{abstract}
On an assigned graph, the problem of Multi-Agent Pathfinding (MAPF) consists in finding paths for multiple agents, avoiding collisions. Finding the minimum-length solution is known to be NP-hard, and computation times grows exponentially with the number of agents. However, in industrial applications, it is important to find feasible, suboptimal solutions, in a time that grows polynomially with the number of agents. Such algorithms exist for undirected and biconnected directed graphs. Our main contribution is to generalize these algorithms to the more general case of strongly connected directed graphs.
In particular, given a MAPF problem with at least two holes, we present an algorithm that checks the problem feasibility in linear time with respect to the number of nodes, and provides a feasible solution in polynomial time.

\end{abstract}
\section{INTRODUCTION}

We consider a graph and a set of agents. Each agent occupies a different node and may move to unoccupied positions.
The Multi-Agent Path Finding (MAPF) problem consists in computing a sequence of movements that repositions all agents to assigned target nodes, avoiding collisions. In this paper, we deal with strongly connected digraphs, directed graphs in which it is possible to reach any node starting from any other node.
The main motivation comes from the management of fleets of automated guided vehicles (AGVs). AGVs move items between different locations in a warehouse. Each AGV follows predefined paths, that connect the locations in which items are stored or processed. We associate the paths' layout to a directed graph.
The nodes represent positions in which items are picked up and delivered,
together with additional locations used for routing. The directed arcs represent the precomputed paths that connect these locations. If various AGVs move in a small scenario, each AGV represents an obstacle for the other ones. In some cases, the fleet can reach a deadlock situation, in which every vehicle is unable to reach its target. Hence, it is important to find a feasible solution to MAPF, even in crowded configurations.

{\bf Literature review.}
Various works address the problem of finding the optimal solution of MAPF (i.e., the solution with the minimum number of moves). For instance, Conflict Based Search (CBS) is a two-level algorithm which uses a search tree, based on conflicts between individual agents (see~\cite{cbs}).
However, finding the optimal solution of MAPF is NP-hard (see~\cite{np}), and computational time grows exponentially with the number of agents.
Search-based suboptimal solvers aim to provide a high quality solution, but are not complete (i.e., they are not always able to return a solution). A prominent example is Hierarchical Cooperative A$^*$ (HCA$^*$) \cite{hca}, in which agents are planned one at a time according to some predefined order.
Instead, rule-based approaches include specific movement rules for different scenarios. They favor completeness at low computational cost over solution quality. Two important rule-based algorithms are TASS \cite{Tass} and \textit{Push and Rotate} \cite{par} \cite{pas}. TASS is a tree-based agent swapping strategy which is complete on every tree, while \textit{Push and Rotate} solves every MAPF instance on graphs that contains at least two holes (i.e., unoccupied vertices). 
Reference~\cite{PMT} presents a method that converts the graph into a tree (as in~\cite{tree}), and solves the resulting problem with TASS. 

The literature cited so far concerns exclusively undirected graphs, where motion is permitted in both directions along graph edges. Fewer results are related to directed graphs. Reference~\cite{MAPF_digraph} proves that finding a feasible solution of MAPF on a general directed graph (digraph) is NP-hard. However, in some special cases this problem can be solved in polynomial time. One relevant reference is~\cite{diBOX}, which solves MAPF on the specific class of biconnected digraphs, i.e., strongly connected digraphs where the undirected graphs obtained by ignoring the edge orientations have no cutting vertices. The proposed algorithm has polynomial complexity with respect to the number of nodes.


{\bf Statement of contribution.}
We consider MAPF on strongly connected digraphs, a class that is more general than biconnected digraphs, already addressed in~\cite{diBOX}.
To our knowledge, this is the first work that considers this specific problem.
Essentially, our approach generalizes the method presented in~\cite{PMT} to digraphs. Namely, we decompose the graph into biconnected components, and use some of the methods presented in~\cite{diBOX} to reconfigure the agents in each biconnected component.
We present a procedure, based on~\cite{tree}, that checks the problem feasibility in linear time with respect to the number of nodes.
Also, we present diSC (digraph Strongly Connected) algorithm that finds a solution for all admissible problems with at least two holes, extending the method in~\cite{PMT}.

\section{PROBLEM DEFINITION}

Let $G=(V,E)$ be a digraph, with vertices $V$ and directed edges $E$.
We assign a unique label to each pebble and hole. Sets $P$ and $H$ contain the labels of the pebbles and, respectively, the holes.
Each vertex of $G$ is occupied by either a pebble or a hole, so that $|V| = |P|+|H|$.
A \textit{configuration} is a function  $\A:P \cup H \rightarrow V$ that assigns the occupied vertex to each pebble or hole. A configuration is \textit{valid} if it is one-to-one (i.e., each vertex is occupied by only one pebble or hole). Set $\C \subset \{P \cup H \to V\}$ represents all valid configurations. 




Given a configuration $\A$ and $u, v \in V$, we denote by $\A[u,v]$ the configuration obtained from $\A$ by exchanging the pebbles (or holes) placed at $u$ and $v$:
\begin{equation} 
	\label{c}
	\A[u,v](q):=  \Bigg\{
	\begin{array}{ll}       	
		
		v, & \text{if } \A(q)  = u ;\\
		
		u, & \text{if } \A(q)  = v ;\\
		
		\A(q), & \text{otherwise } .\\
	\end{array}
\end{equation}
		

Function $\rho: \C \times E \rightarrow \C$ is a partially defined transition function such that $\rho(\A,u \rightarrow v)$ is defined if and only if $v$ is empty (i.e., occupied by a hole). In this case $\rho(\A,u \rightarrow v)$ is the configuration obtained by exchanging the pebble or the hole in $u$ with the hole in $v$. Notation $\rho(\A,u \rightarrow v)!$ means that the function is well-defined. In other words
$\rho(\A, u \rightarrow v)!$ if and only if $(u,v)\in E$ and $A^{-1}(v) \in H$, and, if $\rho(\A, u \rightarrow v)!$,
$\rho(\A, u \rightarrow v)=\A[u,v]$. Note that the hole in $v$ moves along edge $u\rightarrow w$ in reverse direction, while pebble or hole on $u$ moves on $v$.


We represent plans as ordered sequences of directed edges.
It is convenient to view the elements of $E$ as the symbols of a language.
We denote by $E^*$ the Kleene star of $E$, that is the set of
ordered sequences of elements of $E$ with arbitrary length, together with the empty string $\epsilon$:
\[
E^*=\bigcup_{i=1}^\infty E^i \cup \{\epsilon\}.
\]
We extend function $\rho:\C \times E \to \C$ to $\rho: \C \times E^*  \to \C$, by setting $(\forall \A \in \C) \rho(A,\epsilon)!$ and $\rho(\A,\epsilon)=\A$. Moreover,  $(\forall s \in E^*, e \in E, \A \in \C)$  $\rho(\A,  se) !$ if and only if $\rho(\A, s)!$ and $\rho(\rho(\A, s), e)!$ and, if $\rho(\A se)!$, $\rho(\A se)=\rho(\rho(\A s),e)$.
A \emph{move} is an element of $E$, and a \emph{plan} is an element of $E^*$.
Note that $\epsilon$ is the trivial plan that keeps all pebbles and holes on their positions.
We define an equivalence relation $\sim$ on $E^*$, by setting, for $s,t \in E^*$, $s \sim t \leftrightarrow (\forall \A \in \C)\, \rho(\A,s)=\rho(\A,t)$.
In other words, two plans are equivalent if they reconfigure pebbles and holes in the same way.
Given a configuration $\A$ and a plan $f$ such that $\rho(\A, f)!$, a plan $f^{-1}$ is a \emph{reverse} of $f$ if $\A =\rho(\rho(\A, f),f^{-1})$  (i.e., $f^{-1}$ moves each pebble and hole back to their initial positions). We can also write $f f^{-1}\sim \epsilon$, so that $f^{-1}$ behaves like a right-inverse.

	 

Our main problem is the following one:

\begin{defn}{(\textbf{MAPF problem}).}
Given a graph $G = (V,E)$, a pebble set $P$, an initial valid configuration $A^s$, and a final valid configuration $A^t$, find a plan $f$ such that $\A^t(P) = \rho(A^s,f)(P)$.
\end{defn}


If $G$ is an undirected tree, this problem is called \textit{pebble motion on trees} (\textbf{PMT}). A particular case of PMT problem is the \textit{pebble permutation on trees} (PPT), in which the final configuration $\A^t$ is such that $\A^t(P)= \A^s(P)$, that is the final positions are a permutation of the initial ones.

\section{Solving MAPF on undirected graphs}
\label{sec:graphs}
In this section, we recall the planning method for a connected undirected graph presented in~\cite{PMT}. The main idea is to trasform the graph $G=(V,E)$ into a \textit{biconnected component tree} $T:=\T(G)$, and the MAPF problem into a PMT problem. It is possible to prove that the MAPF problem is solvable on $G$ if and only if the corresponding PMT problem is solvable on $T$. Moreover, the solution of MAPF can be obtained from the solution of the corresponding PMT.
                                                                                                                                                                                                                                                                                                                                                                                   
\subsection{Convert MAPF into PMT}

Given a connected graph $G =(V,E)$, we construct the biconnected component tree $\T(G)=(V_T,E_T)$ as follows.
We initialize $V_T = V$, $E_T = E$, and we convert each maximal non-trivial (i.e., with at least three vertices) biconnected component $S=(V_S,E_S) \subset G$ into a star subgraph. The nodes in $V_S$ are the leaves of the star. The internal node of the star is a newly added \textit{trans-shipment vertex}, that play a special role. Indeed, this node cannot host pebbles: pebbles can cross this node, but cannot stop there. More formally, given a trans-shipment vertex $s$, $\rho(\A,(u \rightarrow s)(w \rightarrow v))!$ if and only if $w=s$, $(u,s),(s,v)\in E$, and $\A^{-1}(v) \in H$. If $\rho(\A,(u \rightarrow s)(w \rightarrow v))!$, then $\rho(\A,(u \rightarrow s)(w \rightarrow v))= \A[u,v]$. This means that, if a pebble is moved to a trans-shipment vertex, then it must be immediately moved to another node.

\begin{figure}[h!]
	
	\centering
	\begin{tikzpicture}
		[scale=.5,auto=left,every node/.style={circle,fill=blue!20}]
		\node [scale = 0.6] (n1) at (1,2) {1};
		\node [scale = 0.6] (n2) at (2,3) {2};
		\node [scale = 0.6] (n3) at (3,2) {3};
		\node [scale = 0.6] (n4) at (2,1) {4};
		\node [scale = 0.6] (n5) at (5,2) {5};
		\node [scale = 0.6] (n6) at (7,2) {6};
		\node [scale = 0.6] (n7) at (8,3) {7};
		\node [scale = 0.6] (n8) at (9,2) {8};
		\node [scale = 0.6] (n9) at (8,1) {9};
		\node [scale = 0.6] (n10) at (10,1) {10};
		\node [scale = 0.6] (n11) at (10,3) {11};
		\node [scale = 0.6] (n12) at (12,3) {12};
		\node [scale = 0.6] (n13) at (12,1) {13};
		\foreach \from/\to in {n1/n2,n2/n3,n3/n4,n4/n1,n3/n5,n5/n6,n6/n7,n7/n8,n8/n9,n9/n6,n9/n10,n10/n11,n11/n7,n11/n12,n12/n13,n13/n11}
		\draw (\from) -- (\to);
		
	\end{tikzpicture}


	
	\begin{tikzpicture}
		[scale=.5,auto=left,every node/.style={circle,fill=blue!20}]
		\node [scale = 0.6] (n1) at (1,2) {1};
		\node [scale = 0.6] (n2) at (2.5,3) {2};
		\node [scale = 0.6] (n3) at (4,2) {3};
		\node [scale = 0.6] (n4) at (2.5,1) {4};
		\node [scale = 0.6] (n5) at (5.5,2) {5};
		\node [scale = 0.6] (n6) at (7,2) {6};
		\node [scale = 0.6] (n7) at (8,3) {7};
		\node [scale = 0.6] (n8) at (10.5,2) {11};
		\node [scale = 0.6] (n9) at (8,1) {9};
		\node [scale = 0.6] (n10) at (9.5,1) {10};
		\node [scale = 0.6] (n11) at (9.5,3) {8};
		\node [scale = 0.6] (n15) at (13,3) {12};
		\node [scale = 0.6] (n16) at (13,1) {13};
		\node[rectangle,fill=red!20, scale=0.6] (n12) at (2.5,2) {A};
		\node[rectangle,fill=red!20, scale=0.6] (n13) at (8.75,2) {B};
		\node[rectangle,fill=red!20, scale=0.6] (n14) at (12,2) {C};
		
		\foreach \from/\to in {n12/n1,n12/n2,n12/n3,n12/n4,n3/n5,n5/n6,n6/n13,n13/n7,n13/n8,n13/n9,n13/n10,n13/n11,n8/n14,n14/n15,n14/n16}
		\draw (\from) -- (\to);
		
	\end{tikzpicture}
	
	
	\caption{Undirected graph and corresponding biconnected component tree. $A$, $B$, and $C$ are the trans-shipment vertices.}
	\label{fig:grafo}
\end{figure}
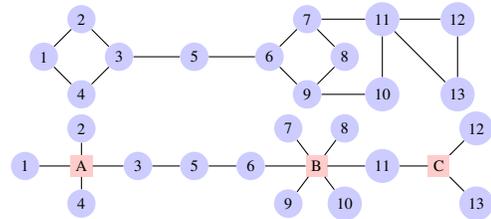

The conversion of $S$ into a star involves the following steps:
\begin{enumerate}
	\item add a trans-shipment vertex $s$,
	\item remove every edge $e\in E_T$,
	\item add the edges $\left\{(u,s)| u \in V_T \right\}$.
\end{enumerate}
Note that $V_T=V \cup \bar{V}$, where $\bar{V}$ is the set of all trans-shipment vertices. $G$ and $\T(G)$ have a similar structure. Biconnected components of $G$ correspond to star subgraphs in $\T(G)$, with trans-shipment vertices as internal nodes. Figure~\ref{fig:grafo} shows an undirected graph and its corresponding biconnected component tree. $G$ and $\T(G)$ have the same number of pebbles and the same number of holes, since trans-shipment vertices are not considered as free. Building $\T(G)$ from $G$ takes a linear time with respect to $|E|$ \cite{BCC}.



Let $\A: P \cup H \rightarrow V$ be a configuration on $G$. We associate it to a configuration on $\T(G)$, $\tilde{\A}: P \cup H \rightarrow V$ such that  $(\forall q \in P \cup H) \,\tilde{\A}(q) = \A(q)$. Note that the codomain of $\tilde{\A}$ is $V$, not $V_T$, since trans-shipment nodes are not present in $G$.
In this way, we associate every MAPF instance on $G$ to a PMT instance on $T$. Reference~\cite{tree} proves the following important result.

\begin{lem}\cite{tree}
	\label{lem:feaspmtmapf}
        Let $G=(V,E)$ be a connected undirected graph, which is not a cycle, and let $\T(G)$ be the corresponding biconnected component tree. Let $\A$ be an initial configuration on $G$ and $\tilde{\A}$ the corresponding configuration on $\T(G)$. Let $a,b\in V$. Then, if $|H|\geq 2$, there is a plan $f_{ab}$ such that $\A[a,b]=\rho(\A,f_{ab})$  if and only if there is a plan $f'_{a'b'}$ such that $\tilde{\A}[a',b']=\rho(\tilde{\A},f'_{a'b'})$.
      \end{lem} 
As a consequence of this Lemma, it follows that:
\begin{theorem}\cite{tree}
	\label{th:iff}
	MAPF on graph $G$ is feasible if and only if PMT on tree $\T(G)$ is feasible.
\end{theorem}
Since feasibility of PMT on a tree $T=(V_T,E_T)$ is decidable in $O(|V_T|)$ time (see~\cite{Auletta}), it follows that:
\begin{theorem}
	\label{feas_graph}
	The feasibility of a MAPF instance on an undirected graph $G=(V_G,E_G)$ is decidable in $O(|V_G|)$ time.
\end{theorem}

\subsection{Solving PMT}
Since solving MAPF is equivalent to solving PMT, we recall the algorithm which solves PMT presented in~\cite{PMT}, inspired by the feasibility test presented in \cite{Auletta}. The idea is to transform PMT into PPT:

\begin{enumerate}
	\item Convert $G$ into the biconnected component tree $\T(G)$ and convert MAPF into PMT;
	\item \textbf{From PMT to PPT}. Reduce the PMT problem to PPT by moving each pebble into one of the target positions (it can be the target position assigned to another pebble). This reduction can be achieved in linear time with respect to $|V_T|$.
	\item  \textbf{Solving PPT instances}.  PPT is solvable if for every pebble $p$ there exists an exchange plan $f_{\bar \A^s(p) \A^t(p)}$, which swaps $p$ with the pebble occupying its target position. Feasibility of the swap between two pebbles can be checked in constant time. We can solve PPT with TASS, proposed in \cite{Tass}. 
	\item Convert the solution of PMT on $\T(G)$ into solutions of MAPF on $G$, using function \textsl{CONVERT-PATH}, presented in detail in \cite{PMT}. 
\end{enumerate}
		


\section{Strongly connected digraphs}

As said, we consider MAPF for \textit{strongly connected digraphs}.

\begin{defn}
A digraph $D=(V,E)$ is \emph{strongly connected} if for each $v,w\in V$, $v\neq w$, there exist a directed path from $v$ to $w$, and a directed path from $w$ to $v$ in $D$.	
\end{defn}

As shown in Proposition~13 of~\cite{feasibility}, in strongly connected digraphs each move is reversible. From this, a more general result follows:

\begin{prop}
  \label{reverse}
  In a strongly connected digraph each plan has a reverse plan.
\end{prop}

Given a digraph $D$, we indicate with $\G(D)$ its \textit{underlying graph},
that is the undirected graph obtained by ignoring the orientations of the edges.
Note that $D$ is strongly connected only if $\G(D)$ is connected. Proposition~\ref{reverse} leads to the following result about the feasibility of MAPF on digraphs:
\begin{theorem}
  \label{thm_strat_rev}
	Let $D=(V_D,E_D)$ be a strongly connected digraph. Then,
	\begin{enumerate}
		\item any MAPF instance on $D$ is feasible if and only if it is feasible on the underlying graph $G=\G(D)$;
		\item feasibility of any MAPF instance on $D$ is decidable in linear time with respect to $|V_D|$.
	\end{enumerate} 
	 
\end{theorem}

\begin{proof}
	\begin{enumerate}
        \item
The necessity is obvious. To prove sufficiency, let $f'$ be a plan which solves a MAPF instance on $\G(D)$. Then we can define a plan $f$ on $D$ in the following way. For each pebble move $u \rightarrow v$ in $f'$, if $(u,v) \in E_D$, we perform move $u \rightarrow v$ on $D$. Otherwise, since $(v,u) \in E_D$, we execute a reverse plan for $v \rightarrow u$, $(v \rightarrow u)^{-1}$, that exists by Proposition~\ref{reverse}. 
		
		\item It follows from Theorem~\ref{feas_graph}. 
	\end{enumerate}
\end{proof}

A direct consequnce of Theorem~\ref{thm_strat_rev} and Theorem~\ref{th:iff} is the following important result:
\begin{cor}
	\label{cor}
	MAPF on strongly connected digraph $D$ is feasible if and only if PMT on tree $\T(\G(D))$ (i.e., the biconnected component tree of the underlying graph of $D$) is feasible.
\end{cor}

The proof of Theorem~\ref{thm_strat_rev} leverages the reversibility of each pebble motion in strongly connected digraphs. It presents a simple algorithm that reduces MAPF for strongly connected digraphs to the undirected graphs case. However, this approach leads to very redundant solutions, since it does not exploit the directed graph structure. This fact is illustrated in Fig.~\ref{bigrafo2}, that shows a digraph $D$ and its associated underlying graph $\G(D)$.

\begin{exam}
	\label{ex}
	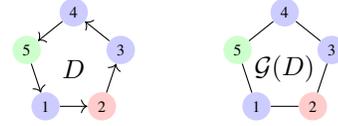
\begin{figure}[h!]
		\centering
		
		\begin{tikzpicture}
			[scale=.5,auto=left]
			\node[circle,fill=blue!20,scale = 0.6] (n1) at (1,0.5) {1};
			\node[circle,fill=red!20,scale = 0.6] (n2) at (2.5,0.5) {2};
			\node[circle,fill=blue!20,scale = 0.6] (n3) at (3,2) {3};
			\node[circle,fill=blue!20,scale = 0.6] (n4) at (1.75,3) {4};
			\node[circle,fill=green!20,scale = 0.6] (n5) at (0.5,2) {5};
		\node at (1.75,1.5) {$D$};
			
			\foreach \from/\to in {n1/n2,n2/n3,n3/n4,n4/n5,n5/n1}
			\draw[->] (\from) -- (\to);

		\end{tikzpicture}	
	\quad \quad \quad 
	\begin{tikzpicture}
		[scale=.5,auto=left]
			\node[circle,fill=blue!20,scale = 0.6] (n1) at (1,0.5) {1};
			\node[circle,fill=red!20,scale = 0.6] (n2) at (2.5,0.5) {2};
			\node[circle,fill=blue!20,scale = 0.6] (n3) at (3,2) {3};
			\node[circle,fill=blue!20,scale = 0.6] (n4) at (1.75,3) {4};
			\node[circle,fill=green!20,scale = 0.6] (n5) at (0.5,2) {5};
		\node at (1.75,1.5) {$\G(D)$};
			
			\foreach \from/\to in {n1/n2,n2/n3,n3/n4,n4/n5,n5/n1}
			\draw[-] (\from) -- (\to);

	\end{tikzpicture}
		\caption{A digraph $D$ and its underlying graph $\G(D)$.}	
		\label{bigrafo2}
	\end{figure}

        A pebble $p$ is placed at node $2$, while all other nodes are free.
        We want to move $p$ to $5$. Plan $f'=(2 \rightarrow 1)(1 \rightarrow 5)$ is a solution of the corresponding problem on $\G(D)$.
        We convert this to a plan on $D$ by applying the method in Theorem~\ref{thm_strat_rev}. Since $(2,1) \not \in E_D$, move $(2 \rightarrow 1)$ is converted into plan $(2,3)(3,4)(4,5)(5,1)$. Similarly, move $(1 \rightarrow 5)$ is converted into $(1,2)(2,3)(3,4)(4,5)$. This solution is redundant, since shorter plan $f=(2 \rightarrow 3)(3 \rightarrow 4)(4 \rightarrow 5)$ solves the overall problem.
      \end{exam}

To find shorter solutions, we avoid using the method described in Theorem~\ref{thm_strat_rev}, and present a method that takes into account the structure of the directed graph. 
In particular, we will exploit the fact that strongly connected digraphs can be decomposed in strongly biconnected components.
In each component, we will use the method presented in~\cite{diBOX}. First, we recall the following definition.

\begin{defn}
	A digraph $D$ is said to be strongly biconnected if $D$ is strongly connected and $\G(D)$ is biconnected.
\end{defn}

We recall that an undirected graph $G$  is biconnected if it is connected and there are no cut vertices, i.e., the graph remains connected after removing any single vertex. The \textit{partially-bidirectional cycle} is a simple example of a strongly biconnected digraph:
\begin{defn}
	A digraph is a \textit{partially-bidirectional cycle} if
	it consists of a simple cycle $C$, plus zero or more edges of
	the type $(u, v)$, where $(v, u) \in C$ (i.e., edges obtained by
	swapping the direction of an edge from $C$).
	
\end{defn}

Reference~\cite{feasibility} shows that strongly biconnected (respectively, strongly connected) digraphs have an open (respectively, closed) ear decompositions.We recall the definitions of open and closed ear decompositions.
Given a graph $D=(V_D,E_D)$ and a sub-digraph $H=(V_H,E_H)$, a path $\pi$ in $D$ is a $H$-path if it is such that its startpoint and its endpoint are in $V_H$, no internal vertex is in $V_H$, and no edge of the path is in $E_{H}$. Moreover, a cycle $C$ in $D$ is a $H$-cycle if it there is exactly one vertex of $C$ in $V_H$.

\begin{defn}
  \label{def_open_closed_ear}
	Let $D =(V_D,E_D)$ be a digraph and $L=[L_0,L_1,\ldots,L_r]$ an ordered sequence of sub-digraphs of $D$, where $L_i=(V_{L_i},E_{L_i})$. We say that $L$ is:
	\begin{enumerate}
		\item a \textit{closed ear decomposition}, if:
	\begin{itemize}
		\item $L_0$ is a cycle,
		\item for all $0 < i \leq r$, $L_i$ is a $D_i$-path or a $D_i$-cycle, where $D_i =(V_{D_i},E_{D_i}) $ with $V_{D_i}=\bigcup_{0\leq j< i}V_{L_j}$ and $E_{D_i}=\bigcup_{0\leq j< i}E_{L_j}$, 
		\item $V_D = \bigcup_{0\leq j\leq r}V_{L_j}$, $E_D = \bigcup_{0\leq j\leq r}E_{L_j}$
		
	\end{itemize}

\item an \textit{open ear decomposition} (\textit{oed}), if it is a closed ear decomposition such that for all $0 < i \leq r$, $L_i$ is a $D_i$-path, (i.e., it is not a $D_i$-cycle).
\end{enumerate}
\end{defn}

In Definition~\ref{def_open_closed_ear}, each $L_i$ is called an \textit{ear}. In particular, $L_0$ is the basic cycle and the other ears are derived ears. An ear is \textit{trivial} if it has only one edge. 

\begin{defn}
We say that an open ear decomposition of a
strongly biconnected digraph is \textit{regular} (\textit{r-oed}) if the basic cycle $L_0$
has three or more vertices, and there exists a non-trivial derived ear with both ends attached to the basic cycle.
\end{defn}
\begin{figure}[h!]
\centering
	
	\begin{tikzpicture}
		[scale=.5,auto=left]
		\node[circle,fill=blue!20,scale = 0.6] (n1) at (2.5,1) {1};
		\node[circle,fill=blue!20,scale = 0.6] (n2) at (4,1) {2};
		\node[circle,fill=blue!20,scale = 0.6] (n3) at (4.5,2.5) {3};
		\node[circle,fill=blue!20,scale = 0.6] (n4) at (3.25,3.5) {4};
		\node[circle,fill=blue!20,scale = 0.6] (n5) at (2,2.5) {5};
		\node[circle,fill=blue!20,scale = 0.6] (n6) at (5.5,3.5) {6};
		\node[circle,fill=blue!20,scale = 0.6] (n7) at (4.25,4.5) {7};
		\node[circle,fill=blue!20,scale = 0.6] (n8) at (0.5,1) {8};
		\node[circle,fill=blue!20,scale = 0.6] (n9) at (0,2.9) {9};
		\node[circle,fill=blue!20,scale = 0.6] (n10) at (1.7,4.5) {10};
		\node at (3.25,2.25) {$L_0$};
		\node at (4.25,3.5) {$L_1$};
		\node at (1.2,3) {$L_2$};
		
		\foreach \from/\to in {n1/n2,n2/n3,n3/n4,n4/n5,n5/n1,n3/n6,n6/n7,n7/n4,n1/n8,n8/n9,n9/n10,n10/n7}
		\draw[->] (\from) -- (\to);

	\end{tikzpicture}
	\caption{Digraph with an open ear decomposition.}	
		\label{bigrafo}
\end{figure}
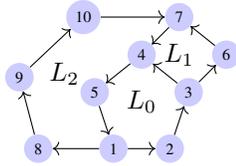                                                                                                                                                         

\begin{obser}
	\label{open}              
	Let $D=(V,E)$ be a digraph with an \textit{oed} $L = [L_0,L_1,\ldots,L_n]$. For each pair $v,w \in V$, there exists a sequence of cycles $C= [C_{1},\ldots,C_{n}]$ such that:
	\begin{itemize}
		\item $v \in V_{C_{1}}$ and $w \in V_{C_{n}}$; 		
		\item for all $j=1,\ldots,n-1$, $\exists a_{j},b_{j} \in V_{C_j} \cap V_{C_{j+1}}$ such that $(a_{j},b_{j} ) \in E$.
	\end{itemize} 

        Figure~\ref{bigrafo} shows a digraph with an \textit{oed} $[L_0,L_1,L_2]$. The sequence of cycles associated to pair $v=2$, $w=10$ is $C = [C_0,C_2]$, where $C_0 = L_0$ and $C_2$ is the subgraph induced by $\{1,8,9,10,7,4,5\}$. Note that  $(4,5) \in C_0 \cap C_2$. The sequence associate to pair $v=1$, $w = 6$ is simply $C = [C_1]$, where $C_1$ is the subgraph induced by $\{1,2,3,6,7,4,5\}$. In fact, nodes $1$ and $6$ belong to the same cycle.
\end{obser}
\begin{proof}
Let $\pi=u_1=v,u_2,\ldots,u_{n-1},u_n=w$ be a shortest path from $v$ to $w$. Let $L_{i}$ be an ear such that $v,u_2\in V_{L_{i}}$. Let $n_{1}$ and $m_{1}$ be the startpoint and endpoint of $L_{i}$. Then, there exists a path $\pi_{1}$ from $m_{1}$ to $n_{1}$ and $C_{1}=\pi_{1} \cup L_{i}$ is the first cycle of the sequence. We initialize $C= [C_1]$ and we set $k=1$. 
Now, if $n>2$, for $j=3,\ldots,n$:
\begin{itemize}
	\item if $u_j \in V_{C_k}$ we go to next iteration;
	\item otherwise, let $\pi_{j-1}$ be a path from $u_j$ to $u_{j-2}$, $C_{k+1}=(V_{C_{k+1}},E_{C_{k+1}})=\pi_{j-1} \cup (u_{j-2},u_{j-1}) \cup (u_{j-1},u_j)$ (note that $u_{j-2},u_{j-1}\in V_{C_{k}} \cap V_{C_{k+1}}$); we add $C_k$ to $C$ and set $k = k+1$, then we go to the next iteration.
\end{itemize} 

\end{proof}

We recall the following results, that characterize strongly biconnected and strongly connected digraphs:
\begin{theorem}
	Let $D$ be a non-trivial digraph.
\begin{itemize}		
	\item  $D$ is strongly biconnected if and only if D has an \textit{oed}. Any cycle can be the starting point of an \textit{oed} \cite{libro16}.
	
    \item $D$ is strongly biconnected if and only if exactly one of the following holds \cite{diBOX}: 
	\begin{enumerate}
\item $D$ is a partially-bidirectional cycle;
		\item $D$ has a \textit{r-oed}.
	\end{enumerate}
\end{itemize}
\end{theorem}

\begin{theorem}\cite{libro3}	
	Let $D$ be a non-trivial digraph. $D$ is strongly connected if and only if D has a closed ear decomposition. 
\end{theorem}                                                                                                                                                                                                                                                                                                                        

\begin{obser}
	\label{graph_composition}
Roughly speaking, this last result means that a strongly connected digraph is composed of non-trivial strongly biconnected components connected by corridors, or articulation points. A \textit{corridor} is a sequence of adjacent vertices $u_1,\ldots,u_n$ such that $(u_i,u_{i+1}), (u_{i+1},u_i)\in E$  for each $i=1,\ldots,n-1$. For example, in Fig.~\ref{digrafo} the subgraph induced by nodes 3,5 and 6 is a corridor. Given a digraph $D=(V,E)$, vertex $v \in V$ is an \textit{articulation point} if its removal increases the number of  connected components of the underlying graph $\G(D)$. In Fig.~\ref{digrafo} nodes 3, 6 and 11 are articulation points.

\end{obser}

\subsection{Solving MAPF on strongly biconnected digraphs}

Reference~\cite{diBOX} shows that all MAPF instances on strongly biconnected digraphs with at least two holes can be solved (or proven to be unsolvable) in polynomial time. 
It also presents Algorithm diBOX, that solves MAPF in the two possible cases of a partially-bidirectional cycle and of a digraph with a \textit{r-oed}.

\textbf{Partially-bidirectional cycle.}
This is the easy case. As no swapping between agents is possible, an instance is solvable if and only if the agents come in the right order in the first place. In this case, only one hole is needed in the digraph. Computing the solution can be performed by diBOX with a time complexity of $O(|V_D|^2)$.

\textbf{Regular open-ear decomposition.}
\label{sec:regularopenear}
This is a more complex case. 
 \begin{prop}
 	\label{oed}
 	~\cite{diBOX} Let $D$ be a strongly biconnected digraph with a \textit{r-oed}, with pebbles $P$ and holes $H$,  with $|H| \geq 2$. For any configurations pair $A^s$, $\A^t$, there exists a plan $f$ such that $\A^t(P) = \rho(A^s,f)(P)$ (i.e., all MAPF instances with at least two holes have a solution).
 \end{prop}

	
	
In particular, diBOX solves any MAPF instance with at least two holes, and finds a solution in $O(|V_D|^3)$ time.

\section{Path planner for strongly connected digraph}
As said, in literature, MAPF has been studied only on connected undirected graphs or on biconnected digraphs. In this section, we consider the more general case of strongly connected digraphs. In particular, we discuss the feasibility of MAPF and present an algorithm (diSC) to find solutions in polynomial time.
We will need some results on the \textit{motion planning problem}. We recall its definition.
\begin{defn}
	\label{def:mp}
	Let $D=(V,E)$ be a digraph, $P$ a set of pebbles. Given a pebble $p\in P$, an initial configuration $\A$, and $v\in V$, the \textit{motion planning problem} (MPP) consists in finding a plan $f$ such that $\bar \A = \rho(\A,f)$ satisfies $\bar \A(p)=v$. We indicate such a plan with notation $\A(p) \Rightarrow v$.
\end{defn}

Reference~\cite{feasibility} discusses the feasibility of the motion planning problem and proves the following:

\begin{theorem}{(Theorem 14 of \cite{feasibility})}
	\label{mp}
	Let $D$ be a strongly biconnected digraph, $P$ a set of pebbles and $H$ a set of holes. Then any MPP on $D$ is feasible if and only if $|H|\geq 1$.
\end{theorem}



For connected undirected graphs, in Section \ref{sec:graphs}, we mentioned that the feasibility of MAPF is decidable in linear time with respect to the number of nodes. Indeed, MAPF can be reduced to PMT. In the following, we show that the same result holds for strongly connected digraphs. In fact, it is possible to define a biconnected component tree $T$ and a corresponding PMT problem such that  MAPF on $D$ is solvable if and only if PMT on $T$ is solvable.

The biconnected component tree of a digraph $D$ is the biconnected component tree $\T(G)$ of the underlying graph $G = \G(D)$. By Theorem~\ref{thm_strat_rev} and Theorem~\ref{th:iff}, it follows that MAPF on $D$ is feasible if and only if the corresponding PMT on $\T(G)$ is feasible.

\begin{figure}[h!]
\centering	

\begin{tikzpicture}
	[scale=.5,auto=left,every node/.style={circle,fill=blue!20}]
	\node [scale=.6] (n1) at (1,2) {1};
	\node [scale=.6] (n2) at (2,3) {2};
	\node [scale=.6] (n3) at (3,2) {3};
	\node [scale=.6] (n4) at (2,1) {4};
	\node [scale=.6] (n5) at (5,2) {5};
	\node [scale=.6] (n6) at (7,2) {6};
	\node [scale=.6] (n7) at (8,3) {7};
	\node [scale=.6] (n8) at (9,2) {8};
	\node [scale=.6] (n9) at (8,1) {9};
	\node [scale=.6] (n10) at (10,1) {10};
	\node [scale=.6] (n11) at (10,3) {11};
	\node [scale=.6] (n12) at (12,3) {12};
	\node [scale=.6] (n13) at (12,1) {13};
	\foreach \from/\to in {n1/n2,n2/n3,n3/n4,n4/n1,n3/n5,n5/n3,n5/n6,n6/n5,n6/n7,n7/n8,n8/n9,n9/n6,n9/n10,n10/n11,n11/n7,n11/n12, n12/n13,n13/n11}
	\draw[->] (\from) -- (\to);

\end{tikzpicture}
\caption{Example of strongly connected digraph: the corresponding underlying graph is shown in Figure~\ref{fig:grafo}.}	
\label{digrafo}

\end{figure}
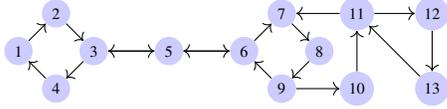

Note that each star subgraph of $\T(\G(D))$ represents a biconnected component of $\G(D)$, which corresponds to a strongly biconnected component of $D$. Indeed, Theorem 9 of \cite{feasibility} defines a one-to-one correspondence between strongly biconnected components of $D$ and biconnected components of $\G(D)$.
We will use the following definition adapted from \cite{tree}:


\begin{defn}
	\label{MP2}
	Let $B = (V,E)$ be a strongly biconnected digraph and $v \not \in V$ be an external node. We consider a digraph $G = (V \cup \{v\},\bar{E})$ with $E \subset \bar{E}$. We say that $G$ is:
	\begin{itemize}
		\item \textit{a strongly biconnected digraph with an entry-attached edge}, if there exists $z\in V$ such that $ \bar{E} = \{(v,z)\} \cup E$ ;
		
		\item \textit{a strongly biconnected digraph with an attached edge}, if there exists $z\in V$ such that $ \bar{E} = \{(v,z),(z,v)\} \cup E$.
		
	\end{itemize}  
	
\end{defn}                                                                                                                                                              

First, we define some basic plans, that we will use to move holes.

\textbf{\textsc{Bring hole from $v$ to $w$.}} Let $\A$ be an initial configuration, such that $v \in \A(H)$ (i.e., $v$ is an unoccupied vertex). Let $\pi=  u_1=w,\ldots,u_n=v$ be a shortest path from $w$ to $v$. 
We define the plan \textsc{Bring hole from $v$ to $w$} as
\begin{equation}
\label{eq:bringhole}
h_{v,w} = (u_{n-1}\rightarrow u_{n},\ldots,u_{1}\rightarrow u_{2}).
\end{equation}
In other words, for each $j$ from $n-1$ to $1$, if there is a pebble on $u_j$, we move it on $u_{j+1}$. 
The new configuration $\bar \A$ is defined as follows:
\begin{equation} 
	\label{cf}                                     
\bar \A(q):=  \Bigg\{
	\begin{array}{lll}       	
		
		u_{j+1}, & \text{if } \A(q)  = u_j &j=1,\ldots,n-1 ;\\
		
		w, & \text{if } \A(q)  = v ; &\\
		
		\A(q), & \text{otherwise} , &\\
	\end{array}
\end{equation}
which means that only pebbles and holes along path $\pi$ change positions.

\textbf{\textsc{Bring back hole from $w$ to $v$.}}
Let $h_{v,w}$ be a plan \textsc{bring hole from $v$ to $w$}. Since the graph is strongly connected, by Proposition \ref{reverse} there exists a reverse plan $h_{v,w}^{-1}$, which returns pebbles and holes to their initial positions. We call \textsc{bring back hole from $w$ to $v$} the plan $h_{v,w}^{-1}$.

\textbf{\textsc{Bring hole from $v$ to a successor of $w$.}} Let $\A$ be an initial configuration, such that $v \in \A(H)$.  Let $\pi=  u_1=w,\ldots,u_n=v$ be a shortest path from $w$ to $v$, where $u_2$ is the successor of $w$ along $\pi$. Then, \textsc{Bring hole from $v$ to a successor of $w$} ($h_{v,s(w)}$) is defined as \textsc{Bring hole from $v$ to $u_2$}.

\textbf{\textsc{Bring back hole from a successor of $w$ to $v$.}}
Let $h_{v,s(w)}$ be a plan \textsc{bring hole from $v$ to $w$}. We call \textsc{bring back hole from $w$ to $v$} its reverse plan $h_{v,s(w)}^{-1}$.
\begin{obser}
	\label{ob:bring}
	Let $\A$ be an initial configuration, $h_{v,w}$ a plan \textsc{bring hole from $v$ to $w$}, and $\bar \A = \rho(\A,h_{v,w})$ the corresponding final configuration. Given $a, b \in V$ with $b \not = v$,  $p=\A^{-1}(a)$ and $q= \A^{-1}(b) $ are the pebbles or holes that occupy $a$, $b$. Then, 	
	$\bar \A[\bar \A(p),\bar \A(q)] =  \overline{\A  [ a , b ]}$. That is, the configuration obtained performing $h_{v,w}$ on $\A[a,b]$ is equal to the one obtained by exchanging $\bar \A(p)$ and $\bar \A(q)$ on $\bar \A$.
	
	
	

\end{obser}

\textbf{\textsc{$k$-Cycle Rotation.}} Let $C=(V_C,E_C)$ be a cycle, with a set of pebbles $P$, and a set of holes $H$, with $|H| \geq 1$. Let $\A$ be an initial configuration, $v \in \A(H)$, and $w \in V_C$ be such that $(v,w) \in E_C$ (i.e., $w$ is the successor of $v$ on $C$). A \textsc{1-Cycle Rotation} over $C$ is defined as \textsc{bring hole from $v$ to $w$}. For $k \in \mathbb{N}$, a \textsc{$k$-Cycle Rotation} over $C$ is obtained by performing $k$ \textsc{1-Cycle Rotations} over $C$.
We denote the plan corresponding to a $k$-Cycle Rotation over $C$ by $r_k^C$.
Let $l_{C}=|V_{C}|$ be the length of cycle $C$. Plan $r_{l_c}^C$ brings all pebbles and holes back to their initial positions. In other words $r_{l_c}^C \sim \epsilon$, where $\epsilon$ is the empty plan. Since
$\epsilon \sim r_{l_C}^C=r_{k}^C r_{l_C-k}^C$, it follows that complementary rotation $r_{l_C-k}^C$ is an inverse plan of $r_k^C$. 

If $C=[C_1,\ldots,C_n]$ is an ordered sequence of cycles, that are subgraphs of the same graph, and $k =(k_1,\ldots,k_n) \in \mathbb{N}^n$, $R_{k}^{C}$ denotes the plan obtained by concatenating a $k_1$-Cycle Rotation over $C_1$, a $k_2$-Cycle Rotation over $C_2$, and analogous rotations over the remaining cycles of $C$, namely:
\[  R_{k}^{C} =r_{k_1}^{C_{1}} \ldots r_{k_n}^{C_{n}}.                           
\] 
Set $s = (s_n,s_{n-1},\ldots,s_1)=(l_{C_n}-k_{n},\ldots,l_{C_1}-k_{1})$ and $\hat C=[C_n,C_{n-1},\ldots,C_1]$. Then
\[
R_k^C R_s^{\hat C}=r_{k_1}^{C_{1}} \ldots r_{k_n}^{C_{n}} r_{s_n}^{C_{n}} \ldots r_{s_1}^{C_{1}} \sim \epsilon,
\]
since $r_{k_n}^{C_{n}} r_{s_n}^{C_{n}} \sim \epsilon$, and analogous reductions holds for the remaining terms. This implies that $R_s^{\hat C}$ is an inverse plan of $R_k^C$. We denote $R_s^{\hat C}$ by $(R_k^C)^+$. In other words, an inverse plan of $R_k^C$ consists in a sequence of complementary rotations, in inverse order.

\begin{obser}
	\label{ob:lemmi}
	In Observation \ref{graph_composition} we noted that a strongly connected digraph $D=(V,E)$ is composed of non-trivial biconnected components and corridors. We can obtain a plan which moves a pebble from $a \in V$ to  $b \in V$ by concatenating four types of movements, that have already been studied in the case of an undirect connected graph in~\cite{tree}. With the following Lemmas we adapt the results of~\cite{tree} to the more general case of a strongly connected directed graph:
	\begin{enumerate}
		\item within the same strongly biconnected component: \textit{Stay in Lemma} \ref{stay};
		\item from a corridor to a strongly biconnected component (or viceversa): \textit{Attached-Edge Lemma} \ref{cycle};
		\item from a strongly biconnected component to another one, connected by an articulation point: \textit{Two Biconnected Components Lemma} \ref{twoedge};
		\item from a node to another one of the same corridor.
	\end{enumerate}
\end{obser}
                                                                                                                                                                                                                                                                                                                                                                                                                                                         
First, we prove the following Lemma, which will be useful for the other results.
\begin{lem}{\textbf{Entry Lemma.}}
	\label{entry}
	Let $P$ be a set of pebbles and $H$, with $|H| \geq 2$, a set of holes on $G= (V \cup \{v\},\bar{E})$, where $G$ is a strongly biconnected digraph with an entry-attached edge $(v,y)$ (see Definition \ref{MP2}). Let $\A$ be a configuration, $p\in P$ such that $\A(p)=v $, and $w \in \A(H)$. Let $ \A[v,w]$ be the configuration defined in (\ref{c}). Then, there exists a plan $f_{v w}$ such that $ \A[v,w]= \rho(\A,f_{vw})$, i.e., that moves $p$ from $v$ to $w$, without altering the locations of the other pebbles. In particular, we can write this plan as
        \[f_{v w} = h_{h s(w)}R_{k}^{C} (v\rightarrow y) (R_{k}^{C})^+ vh_{h s(w)}^{-1},\]
where $C=[C_1,\ldots,C_n]$ is a sequence of cycles, $h$ a hole, and $k \in \mathbb{N}^n$.
\end{lem}
\begin{proof}
	Fig.~\ref{fig:entry} illustrates this proof.
	Let $y\in V$ be such that $(v,y)\in E$. 
	Let $h_1$ be the hole in $w$ ($\A(h_1)=w$), and let $h_2$ be another hole in $V\setminus\{w\}$ (note that $h_2$ exists since we are assuming that $|H|\geq 2$).
	If $G$ is a partially-bidirectional cycle, we set $n=1$ and $C_1=C_G$, where $C_G$ is the directed cycle contained in $G$. 
        We perform a \textsc{$d(w,y)$-Cycle Rotation} over cycle $C_G$, where $d(w,y)$ is the distance between nodes $w$ and $y$.
In this way, hole $h_1$ moves to $y$. Next, pebble $p$ moves on $y$ with $v \rightarrow y$. Finally, we perform the complementary \textsc{$L_{C_G}-d(y,w)$-Cycle Rotation} over $C_G$, in order to move $p$ to $w$. Namely, the plan is $f_{v w} = r_{d(w,y)}^{C_G} (v\rightarrow y) (r_{d(w,y)}^{C_G})^{+} $, and the final configuration is $\bar \A[v,w]$. Indeed, apart from $p$ and $h_1$, which are exchanged, all pebbles and holes are moved $l_{C_G}$ times, which means that they complete a full revolution, returning to their initial positions. 	
If $G$ has a \textit{r-oed} $L=[L_0,\ldots,L_r]$, let $u$ be a successor of $w$ such that $\A(h_2)=u$. If it does not exist, we perform \textsc{bring hole from $\A(h_2)$ to a successor of $w$} and set $u$ as the successor of $w$, which corresponds to the new position of $h_2$. At the end, we will bring back $h_2$ to its initial position with \textsc{bring back hole from a successor of $w$ to $\A(h_2)$}. By Observation \ref{open}, there exists a sequence of cycles $C= [C_{i_1},C_{i_2},\ldots,C_{i_n}]$ such that $v \in C_{i_1}$ and $u \in C_{i_n}$ and for all $j=1,\ldots,n$, $C_{i_j} \cap C_{i_{j+1}}$ has at least two nodes $a_{i_j}$ and $b_{i_j}$  with $(a_{i_j},b_{i_j} ) \in E$. Starting from $C_{i_1}$, we perform the following operations: we perform a \textsc{$d(w,a_{i_1})$-Cycle Rotation} over cycle $C_{i_1}$; for all $j=2,\ldots,n-1$ we perform a \textsc{$d(a_{i_j},a_{i_{j+1}})$-Cycle Rotation} over cycle $C_{i_j}$; we perform a \textsc{$d(a_{i_{n-1}},y)$-Cycle Rotation} over cycle $C_{i_n}$. Setting $k=(d(w,a_{i_1}),d(a_{i_2},a_{i_{3}}),\ldots,d(a_{i_{n-1}},y))$, this sequence of rotations corresponds to $R_k^C$.
Then, we move $p$ to $y$ and perform the inverse sequence $(R_k^C)^+$.
At the end of this plan, $p$ is on $w$ and all other pebbles are in their initial positions. Hence, the overall plan that allows us to prove the thesis is $ f_{vw} = h_{h_2 s(w)}R_k^C (v \rightarrow y) (R_k^C)^+ h_{h s(w)}^{-1}$. 

\end{proof}
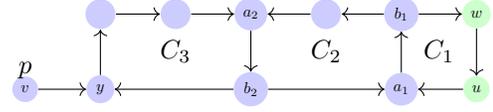
\begin{figure}[h!]
\centering
	
	\begin{tikzpicture}
		[scale=.5,auto=left]
		\node[circle,fill=blue!20,scale=.6] (n1) at (0,1) {$v$};
		\node[circle,fill=blue!20,scale=.6] (n2) at (2,1) {$y$};
		\node[circle,fill=blue!20,scale =1.2] (n3) at (2,3) {};
		\node[circle,fill=blue!20,scale =1.2] (n4) at (4,3) {};
		\node[circle,fill=blue!20,scale=.6] (n5) at (6,3) {$a_2$};
		\node[circle,fill=blue!20,scale=.6] (n6) at (6,1) {$b_2$};
		\node[circle,fill=blue!20,scale=.6] (n7) at (10,1) {$a_1$};
		\node[circle,fill=blue!20,scale=.6] (n8) at (10,3) {$b_1$};
		\node[circle,fill=blue!20,scale =1.2] (n9) at (8,3) {};
		\node[circle,fill=green!20,scale=.6] (n10) at (12,3) {$w$};
		\node[circle,fill=green!20,scale=.6] (n11) at (12,1) {$u$};
		\node at (4,2) {$C_3$};
		\node at (8,2) {$C_2$};
		\node at (11,2) {$C_1$};
		\node at (0,1.5) {$p$};
		\foreach \from/\to in {n1/n2,n2/n3,n3/n4,n4/n5,n5/n6,n6/n2,n6/n7,n7/n8,n8/n9,n9/n5,n8/n10,n10/n11,n11/n7}
		\draw[->] (\from) -- (\to);

	\end{tikzpicture}
	\caption{$f_{vw} = r_2^{C_1}\, r_3^{C_2} \, r_2^{C_3}\, (v \rightarrow y)\,  r_3^{C_3}\, r_2^{C_2}\, r_2^{C_1}$.}
	\label{fig:entry}	
\end{figure}

\begin{lem}{\textbf{Stay in Lemma.}}
	\label{stay}
	Let $P$ be a set of pebbles and $H$, with $|H| \geq 2$, be a set of holes on $D=(V ,E)$, a strongly biconnected digraph with a \textit{r-oed}.  Let $\A$ be a configuration, $p \in P$ such that $\A(p)=v $, and $w \in \A(H)$. Let $ \A[v,w]$ be a configuration defined as in (\ref{c}), then there exists a plan $f_{v w}$ such that $\A[v,w]= \rho(\A,f_{vw})$.
\end{lem}

\begin{proof}
	This is a direct consequence of Proposition \ref{oed}.
	\end{proof}

\begin{lem}{\textbf{Attached-Edge}}
	\label{cycle}
		Let $P$ be a set of pebbles and $H$ a set of holes on $D=(V \cup \{v\}, \bar{E})$, a strongly biconnected digraph with an attached edge such that $|H| \geq 2$.  Let $\A$ be a configuration, $p \in P$ such that $\A(p)=u $, and $w \in \A(H)$. Let $ \A[u,w]$ be a final configuration defined as in (\ref{c}), then there exists a plan $f_{u w}$ such that $ \A[u,w]= \rho(\A,f_{uw})$.
		 
\end{lem}

\begin{figure}[h!]
\begin{subfigure}[t]{0.22\textwidth}
		
	\begin{tikzpicture}
		[scale=.5,auto=left]
		\node[circle,fill=blue!20,scale =0.6] (n1) at (2,1) {$p_1$};
		\node[circle,fill=blue!20,scale = 0.6] (n2) at (1,2) {$p_2$};
		\node[circle,fill=green!20,scale = 1.2] (n3) at (2,3) {};
		\node[circle,fill=blue!20,scale =0.6] (n4) at (3.5,3) {$p_3$};
		\node[circle,fill=blue!20,scale = 0.6] (n5) at (4.5,2) {$p_4$};
		\node[circle,fill=red!20,scale = 0.6] (n6) at (3.5,1) {$p$};
		\node[circle,fill=green!20,scale = 1.2] (n7) at (6.3,2) {};
		
		\node[scale = 0.8] at (4.2,1) {$u$};
		\node[scale = 0.8] at (5,2.5) {$y$};
		\node[scale = 0.8] at (1.4,3) {$w$};
		\node[scale = 0.8] at (7,2) {$v$};
		\foreach \from/\to in {n1/n2,n2/n3,n3/n4,n4/n5,n5/n6,n6/n1,n5/n7,n7/n5}
		\draw[->] (\from) -- (\to);

	\end{tikzpicture}
\caption{Initial positions.}
\end{subfigure}
\begin{subfigure}[t]{0.22\textwidth}
		
	\begin{tikzpicture}
		[scale=.5,auto=left]
		\node[circle,fill=blue!20,scale =0.6] (n1) at (2,1) {$p_2$};
		\node[circle,fill=green!20,scale = 1.2] (n2) at (1,2) {};
		\node[circle,fill=blue!20,scale = 0.6] (n3) at (2,3) {$p_3$};
		\node[circle,fill=blue!20,scale =0.6] (n4) at (3.5,3) {$p_4$};
		\node[circle,fill=red!20,scale = 0.6] (n5) at (4.5,2) {$p$};
		\node[circle,fill=blue!20,scale = 0.6] (n6) at (3.5,1) {$p_1$};
		\node[circle,fill=green!20,scale = 1.2] (n7) at (6.3,2) {};

		\foreach \from/\to in {n1/n2,n2/n3,n3/n4,n4/n5,n5/n6,n6/n1,n5/n7,n7/n5}
		\draw[->] (\from) -- (\to);

	\end{tikzpicture}
\caption{$r_{d(u,y)}$.}
\end{subfigure}	

\vspace{4mm}

\begin{subfigure}[t]{0.22\textwidth}
	\centering
	\begin{tikzpicture}
		[scale=.5,auto=left]
		\node[circle,fill=blue!20,scale =0.6] (n1) at (2,1) {$p_4$};
		\node[circle,fill=green!20,scale = 1.2] (n2) at (1,2) {};
		\node[circle,fill=blue!20,scale = 0.6] (n3) at (2,3) {$p_1$};
		\node[circle,fill=blue!20,scale =0.6] (n4) at (3.5,3) {$p_2$};
		\node[circle,fill=green!20,scale = 1.2] (n5) at (4.5,2) {};
		\node[circle,fill=blue!20,scale = 0.6] (n6) at (3.5,1) {$p_3$};
		\node[circle,fill=red!20,scale = 0.6] (n7) at (6.3,2) {$p$};

		\foreach \from/\to in {n1/n2,n2/n3,n3/n4,n4/n5,n5/n6,n6/n1,n5/n7,n7/n5}
		\draw[->] (\from) -- (\to);

	\end{tikzpicture}
\caption{($y \rightarrow v$) $r_k$.}
\end{subfigure}
\begin{subfigure}[t]{0.22\textwidth}
	\centering
	\begin{tikzpicture}
		[scale=.5,auto=left]
		\node[circle,fill=blue!20,scale =0.6] (n1) at (2,1) {$p_1$};
		\node[circle,fill=blue!20,scale = 0.6] (n2) at (1,2) {$p_2$};
		\node[circle,fill=red!20,scale = 0.6] (n3) at (2,3) {$p$};
		\node[circle,fill=blue!20,scale =0.6] (n4) at (3.5,3) {$p_3$};
		\node[circle,fill=blue!20,scale = 0.6] (n5) at (4.5,2) {$p_4$};
		\node[circle,fill=green!20,scale = 1.2] (n6) at (3.5,1) {};
		\node[circle,fill=green!20,scale = 1.2] (n7) at (6.3,2) {};

		\foreach \from/\to in {n1/n2,n2/n3,n3/n4,n4/n5,n5/n6,n6/n1,n5/n7,n7/n5}
		\draw[->] (\from) -- (\to);

	\end{tikzpicture}
	\caption{($v \rightarrow y$) $r_{d(y,w)}$.}
\end{subfigure}
\vspace{3mm}
\caption{Cycle with an attached-edge.}	
\label{fig:att}
\end{figure}
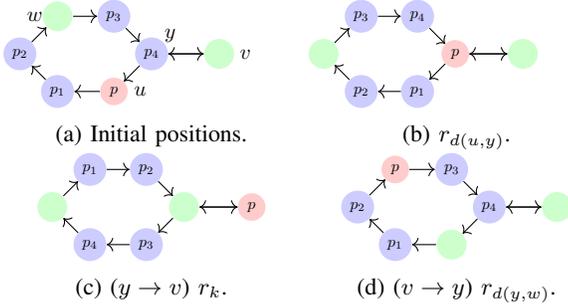

\begin{proof}
	If the strongly biconnected component of $D$ has a \textit{r-oed}, the proof follows from Lemma \ref{stay}. Otherwise, the strongly biconnected component is a partially-bidirectional cycle $G=(V,E)$. We consider the cycle $C_G$ contained in $G$, which has lenght $l_{C_G}$. Let $h_1$ be the hole on $w$ and $h_2$ another hole (which exists, since $|H|\geq2$). Without loss of generality, suppose that $\A(h_2)=v$. Indeed, if this were not the case, we can \textsc{bring hole from $\A(h_2)$ to $v$} and finally bring back it to its initial position. In this case, the new initial configuration is $\bar A= \rho(\A,h_{\A(h_2) v})$ and we have to consider $\bar u = \A(p)$ and $\bar w = \bar \A(h_1)$. Let $y \in V$ be the cycle node that shares and arc with $v$, and consider the distances from $u$ and $w$ to $y$, $d_1 := d(u,y)$ and $d_2 := d(w,y)$. Performing a \textsc{$d_1$-Cycle Rotation} over $C_G$, we move $p$ from $w$ to $y$. Then, we move $p$ on $v$ with $y \rightarrow v$. Now, let
	\[ k =  \begin{cases}
		 d_2-d_1 & if\; d_2\geq d_1,\\
		 
		l+d_2-d_1 & if \; d_2<d_1.
		\end{cases} \]
	We perform a \textsc{$k$-Cycle Rotation} over $C_G$, so that we move the hole $h_1$ from  $w$ to $y$.  Next, we move $p$ from $v$ to $y$ with $v \rightarrow y$. Finally, we perform a \textsc{$d(y,w)$-Cycle Rotation} over $C_G$ to move $p$ on $w$. To conclude, $f_{uw} = r_{d(u,y)}^{C_G} \,(y \rightarrow v)\, r_{k}^{C_G}\, (v \rightarrow y)\, r_{d(y,w)}^{C_G}$. Fig. \ref{fig:att} illustrates this proof.
	
\end{proof}

Next lemma deals with the case of two biconnected components joined by an articulation point like, e.g., $\{6,7,8,9,10,11\}$ and $\{11,12,13\}$ in Figure \ref{digrafo}, where the articulation point is node 11.

\begin{lem}{\textbf{Two Biconnected Components.}}
	\label{twoedge}
		Let $P$ be a set of pebbles and $H$, with $|H|\geq 2$, a set of holes on $D=(V , E)$, a strongly connected digraph, composed of two biconnected components joined by an articulation point.  Let $\A$ be a configuration, $p \in P$ be such that $\A(p)=a $, and $b \in \A(H)$. Let $\A[a,b]$ be a final configuration defined as in (\ref{c}). Then, there exists a plan $f_{a b}$ such that $\A[a,b]= \rho(\A,f_{ab})$.
		  
\end{lem}

\begin{proof} Let  $B_1=(V_1,E_1)$ and $B_2=(V_2,E_2)$ be the two biconnected components and $v$ be the articulation point, i.e., $V_1 \cup V_2 = V$, $E_1 \cup E_2 = E$, $V_1 \cap V_2 = \{v\}$,   and $E_1 \cap E_2 = \emptyset$. 
Let $h_1$ be the hole on $b$, and let $h_2$ be another hole. We discuss different cases.

\begin{enumerate}
	
 \item  $a \in V_1\setminus\{v\}$ and $b\in V_2\setminus\{v\}$.
 
Without loss of generality, we assume that $\A(h_2)=v$. Indeed, if this is not the case, we can \textsc{bring hole from $\A(h_2)$ to $v$}, and the new initial configuration is $\bar \A = \rho(\A,h_{\A(h_2) v})$. Note that $h_{\A(h_2) v}$ could change the position either of $h_1$ or of $p$, i.e., either $\bar a = \bar \A(p)\neq a$ or $\bar b = \bar \A(h_1)\neq b$. By the procedure described below, we will reach  $\bar \A[\bar a, \bar b]$, and at that point we will need to perform \textsc{bring back hole from $v$ to $\A(h_2)$} in order to obtain, by Observation \ref{ob:bring}, $\A [a,b]=\rho(\bar \A[\bar a, \bar b], h_{v, \A(h_2) }^{-1})$.

Assuming $\A(h_2)=v$, let $y \in V_1$ be such that $(y,v)\in E_1$, i.e., $y$ is a predecessor of $v$.
   Now, by \textit{Entry Lemma}~\ref{entry}, there exists a plan $g$ such that, setting $\A^1 = \rho(\A,g)$, $\A^1(p) = b$ and, for all $q \in H \cup P \setminus \{p\}$ such that $\A^1(q) \in V_2$, $\A^1(q)=\A(q)$ holds. Plan $g$ is defined as follows

   
   \[ g = h_{v,s(b)}R_k^C t_{a,v} (R_k^{C})^+ h_{v,s(b)}^{-1},  \] 
   
   where $C$ is a sequence of cycles, $k$ a vector, and $t_{a,v}$ is a plan which moves pebble $p$ to the unoccupied vertex $v$. In particular, if $B_1$:
   
   \begin{itemize}
   	\item has a \textit{r-oed}: $t_{a,v} = a \Rightarrow v$, (see Definition \ref{def:mp}) which by Theorem \ref{mp} exists since in $B_1$ there is at least one hole ($h_1$);
   	\item is a partially-bidirectional cycle:  $t_{a,v} = r_{d(a,v)}^{C_{B_1}}$.                              
   \end{itemize}

   After performing $g$, both holes $h_1$ and $h_2$ are in $B_1$ (in particular, $\A^1(h_2)=v$). If $B_1$ has a \textit{r-oed}, by Lemma~\ref{stay} there exists a plan $f$ so that $\A^2 = \rho(\A^1,f)$ is such that $\A^2(h_1) = a$ and for all $q \in P \cup H \setminus \{h_1\} $, $\A^2(q) \in V_1$, $\A^2(q)=\A(q)$. So, finally $\A^2 = \A[a,b]$. If $B_1$ is a partially-bidirectional cycle, performing $r_{d(v,a)}^{C_{B_1}}$ is sufficient to bring $p$ on $a$ and the other pebbles of $B_1$ on their initial positions.

 \item Suppose that $a,b \in V_1$. 
\begin{itemize}
	\item $B_1$ has a \textit{r-oed}. We can assume without loss of generality that $\A(h_2) \in V_1$ (if not, it would be enough to  \textsc{bring hole from $\A(h_2)$ to $v$} and finally \textsc{bring back hole from $v$ to $\A(h_2)$}). Since in $B_1$ there are two holes, by Lemma \ref{stay} we can move $p$ to $b$ without changing the final position of the other pebbles. 
	
	\item $B_1$ is a cycle. We can assume without loss of generality that $\A(h_2) \in V_2 \setminus \{v\}$. Then, by point $1)$ of this proof, we can first move $p$ from $a$ to $\A(h_2)$ and then from $\A(h_2)$ to $b$, without changing the final position of the other pebbles.
\end{itemize}

\end{enumerate}
\end{proof}

These results allow us to prove that feasibility of a MAPF on a strongly connected digraph with at least two holes is equivalent to feasibility of the corresponding PMT. This is a consequence of the following theorem.
\begin{theorem}\label{th:feasibility}
	Let $P$ be a set of pebbles and $H$ a set of holes on a strongly connected digraph $D=(V,E)$, which is not a partially-bidirectional cycle, and let $T=(V_T,E_T)$ be the corresponding biconnected component tree. Let $\A$ be an initial configuration on $D$ and $\tilde{\A}$ the corresponding configuration on $T$. Let $a,b\in V$ and $p\in P$ be a pebble on $a$. Then, if $|H|\geq 2$, there is a plan $f_{ab}$ on $D$ such that $\A[a,b]=\rho(\A,f_{ab})$ if and only if there is a plan $f'_{ab}$ on $T$ such that $\tilde{\A}[a,b]=\rho(\tilde{\A},f'_{a'b'})$.
\end{theorem}
\begin{proof}	
	Let $f_{ab}$ be a plan on $D$. For each single move $m = u \rightarrow v$ in this plan, recalling Observation \ref{graph_composition} and noting that $u,v$ either belong to the same biconnected component or to the same corridor, there are two cases. In the first case there exists a non-trivial strongly biconnected component $B=(V_B,E_B)$ such that $u,v \in V_B$. Then, we define $m'$, a corresponding plan on $T$, as follows: let $S$ be the star in $T$ corresponding to $B$, and let $s$ be the trans-shipment vertex of $S$; then, $m'=(u \rightarrow s) (s \rightarrow v)$. In the second case, i.e.,  $u,v$ belong to the same corridor, then $m'=m$. $f'_{ab}$ is defined as the composition of all the moves $m'$ just described.
	
	
Conversely, let $f'_{ab}$ be a plan on $T$. Then, by Observation \ref{ob:lemmi}  
	\begin{enumerate}
	\item If $a$ and $b$ are not in the same star on $T$, then they are not in the same strongly biconnected component on $D$. By Observation \ref{ob:lemmi} $f_{ab}$ will be composed by movements: from/to a corridor to/from a strongly biconnected component ($f_{ab}$  exists by \textit{Attached-Edge Lemma} \ref{cycle});
	 from a strongly biconnected component to another one, connected by an articulation point ($f_{ab}$ exists by \textit{Two Biconnected Components Lemma} \ref{twoedge});
	from a node to another one of the same corridor ($f'_{ab}=f_{ab}$).

	\item If $a$ and $b$ belong to the same "star" on $T$, then $a$ and $b$ belong to the same strongly biconnected component $B_1=(W_1,F_1)$ on $D$. There are two possibilities:
	\begin{enumerate}
		\item $B_1$ has at least two holes. In this case, if $B_1$ has a \textit{r-oed}, by Lemma \ref{stay} $f_{a,b}$ exists. If $B_1$ is a partially-bidirectional cycle there are two cases: 
		\begin{itemize}
			\item there is another biconnected component $B_2=(W_2,F_2)$ such that $B_1$ and $B_2$ are joined by an articulation point. In this case existence of $f_{a,b}$ follows from Lemma \ref{twoedge};
			\item there is a node $v \in V\setminus W_1$ such that a node $w\in W_1$ with $(v,w),(w,v)\in E$ exists. Therefore, $G = (W_1 \cup \{v\},F_1  \cup \{(v,w),(w,v)\}) $ is a cycle with an attached edge, and 
 existence of $f_{a,b}$ follows from Lemma \ref{cycle};
		\end{itemize}
		\item $B_1$ has only one hole. In this case, let $h_1$ be the hole on $b$ and $h_2$ a hole such that $\A(h_2) = w \not \in W_1$ (which exists since $|H|\geq 2$) and $u\in W_1$ a node different from $a$ and $b$ (which exists since non-trivial biconnected components have at least three nodes). Let $h_{w,u}$ be the plan \textsc{bring hole from $w$ to $u$} which moves the hole, and $\bar \A =(\A,h_{w,u})$. Then:
		
		\begin{enumerate}
			\item[a')] if $\bar \A(p),\bar \A(h_1) \in W_1$, then we do perform \textsc{bring hole from $w$ to $u$}, we replace $a$ and $b$ with $\bar \A(p)$ and $\bar \A(h_1)$, and by point $a)$ we find the plan $f_{\bar \A(p)\bar \A(h_1)}$; finally we perform $h_{w,u}^{-1}$ and $h_2$ returns to its initial position.
			
			\item[b')] if $\bar \A(h_1) \in W_1$ but $\bar \A(p) \not \in W_1$, we do perform \textsc{bring hole from $w$ to $u$} and we fall into the case where start and final position of the pebble are not in the same biconnected component. Therefore, $f_{ab}=h_{w,u}f_{\bar \A(p)\bar \A(h_1)}h_{w,u}^{-1}$, where $f_{\bar \A(p)\bar \A(h_1)}$ exists by point $1)$.
			
			\item[c')] if $\bar \A(p) \in W_1$ but $\bar \A(h_1) \not \in W_1$, we do not perform \textsc{bring hole from $w$ to $u$}. First, we move $h_1$ away from $b$ by \textsc{bring hole from $b$ to $u$}.
Then, we perform \textsc{bring hole from $w$ to $b$}.
Then, we can proceed as in a') or b') with $u$ replaced by $b$. Finally, we will need to perform \textsc{bring back hole from $u$ to $b$}. In formulas, given $\tilde{\A}= \rho(\A,h_{b,u}h_{w,b})$ the final plan is $f_{ab}=h_{b,u}h_{w,b}f_{\tilde{\A}(p)\tilde{\A}(h_1)} h_{w,b}^{-1}h_{b,u}^{-1} $.
		\end{enumerate} 
	\end{enumerate} 

\end{enumerate}

\end{proof}




\subsection{Algorithm diSC}

The idea of this algorithm is to use the same strategy to solve MAPF on undirected graphs, described in Section \ref{sec:graphs}. The main steps are the following ones:
 \begin{enumerate}
 	\item Convert the digraph $D$ into a tree $T$ and consider the corresponding PMT problem.
 	\item Convert the PMT problem into the PPT problem and solve it.
 	\item Convert solution plans on $T$ into plans on $D$ by a function \textit{CONVERT-PATH}, which is based on Theorem~\ref{th:feasibility}.
 \end{enumerate}


\begin{theorem}
	diSC finds the solution of a MAPF instance with at least two holes on $D=(V,E)$ in polynomial time with respect to $|V|$.
\label{th:pol}
\end{theorem}

\begin{proof}
	Convert $D$ into $\T(D)$ takes $O(|V|)$ time \cite{BCC}. Solve PMT takes polynomial time with respect to $|V|$ \cite{Tass}. \textit{CONVERT-PATH} uses diBOX to move pebbles within a strongly biconnected component of the digraph, which takes $O(|V|^3)$ time \cite{diBOX}.  
\end{proof}

	
	

	
 
 \section{Experimental Results}                                                                                                                                                                     

 We implemented the diSC algorithm in Matlab. To evaluate its behaviour, we generated random graphs with a number of nodes that ranges from 20 to 100, with increments of 5 nodes. For every number of nodes, we generated a set of 200 graphs. 
In order to generate test graphs with multiple biconnected components, we used the following procedure.
First, we create a random connected undirected graph with function \textit{networkx.connected\_watts\_strogatz\_graph()}, contained in the Networkx library\footnote{https://networkx.org/}. Then, we construct a maximum spanning rooted tree.
We process the tree nodes with a breadth-first order. Every node that has a number $n$ of children higher than $1$ is converted into a biconnected component, together with its children, with the following method.
We substitute the parent node and its children with a directed cycle with a random number of nodes lower or equal than $n+1$. Then, we add directed ears of random length (but sufficiently small, not to exceed the total number of $n+1$ nodes assigned to the biconnected component) and random initial and final nodes, until the number of nodes in the resulting biconnected component equals $n+1$.
After processing the tree, every remaining undirected edge $\{u,v\}$ is converted into two directed edges $(u,v)$, $(v,u)$.
\begin{figure}[h!]
\begin{minipage}[t]{0.22\textwidth}
\centering
	
	\begin{tikzpicture}
		[scale=.65,auto=left]
		\node[circle,fill=blue!20, scale = 0.6] (n1) at (2,2) {1};
		\node[circle,fill=blue!20, scale = 0.6] (n2) at (1,1) {2};
		\node[circle,fill=blue!20, scale = 0.6] (n3) at (2,1) {3};
		\node[circle,fill=blue!20, scale = 0.6] (n4) at (3,1) {4};
		\node[circle,fill=blue!20, scale = 0.6] (n5) at (0,0) {5};
		\node[circle,fill=blue!20, scale = 0.6] (n6) at (1,0) {6};
		\node[circle,fill=blue!20, scale = 0.6] (n7) at (2,0) {7};
		\node[circle,fill=blue!20, scale = 0.6] (n8) at (3,0) {8};
		\foreach \from/\to in {n1/n2,n1/n3,n1/n4,n2/n5,n2/n6,n2/n7,n4/n8}
		\draw[-] (\from) -- (\to);

	\end{tikzpicture}
\end{minipage}
\begin{minipage}[t]{0.22\textwidth}
\centering
	
	\begin{tikzpicture}
		[scale=.65,auto=left]
		\node[circle,fill=blue!20, scale = 0.6] (n1) at (2,2) {1};
		\node[circle,fill=blue!20, scale = 0.6] (n2) at (1,1) {2};
		\node[circle,fill=blue!20, scale = 0.6] (n3) at (2,0) {3};
		\node[circle,fill=blue!20, scale = 0.6] (n4) at (3,1) {4};
		\node[circle,fill=blue!20, scale = 0.6] (n5) at (0,1) {5};
		\node[circle,fill=blue!20, scale = 0.6] (n6) at (1,0) {6};
		\node[circle,fill=blue!20, scale = 0.6] (n7) at (0,0) {7};
		\node[circle,fill=blue!20, scale = 0.6] (n8) at (4,1) {8};
		\foreach \from/\to in {n1/n2,n2/n3,n2/n5,n3/n4,n4/n1,n4/n8,n5/n6,n5/n7,n6/n2,n7/n6,n8/n4}
		\draw[->] (\from) -- (\to);

	\end{tikzpicture}
\end{minipage}
\caption{From tree to strongly connected digraph.}
\end{figure}
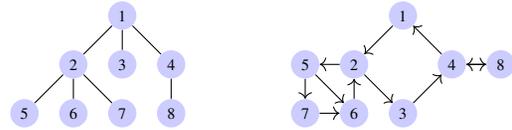

First, we ran the algorithm varying the number of nodes: for every generated graph (200 for every different number of nodes), we created a MAPF problem instance, with 10 agents and random initial and final positions. Then, we ran the algorithm on the set of 200 40-nodes random graphs, with a number of agents varying from 1 to 14. We used a \textit{Intel(R) Core(TM) i7-4510U CPU @ 2.60 GHz} processor with 16 GB of RAM.
For each obtained solution, we recorded the overall number of moves and the computation time. 
\begin{figure}[h]
\begin{minipage}[t]{0.5\textwidth}
\centering
\includegraphics[width=0.49\textwidth, height=2.7cm]{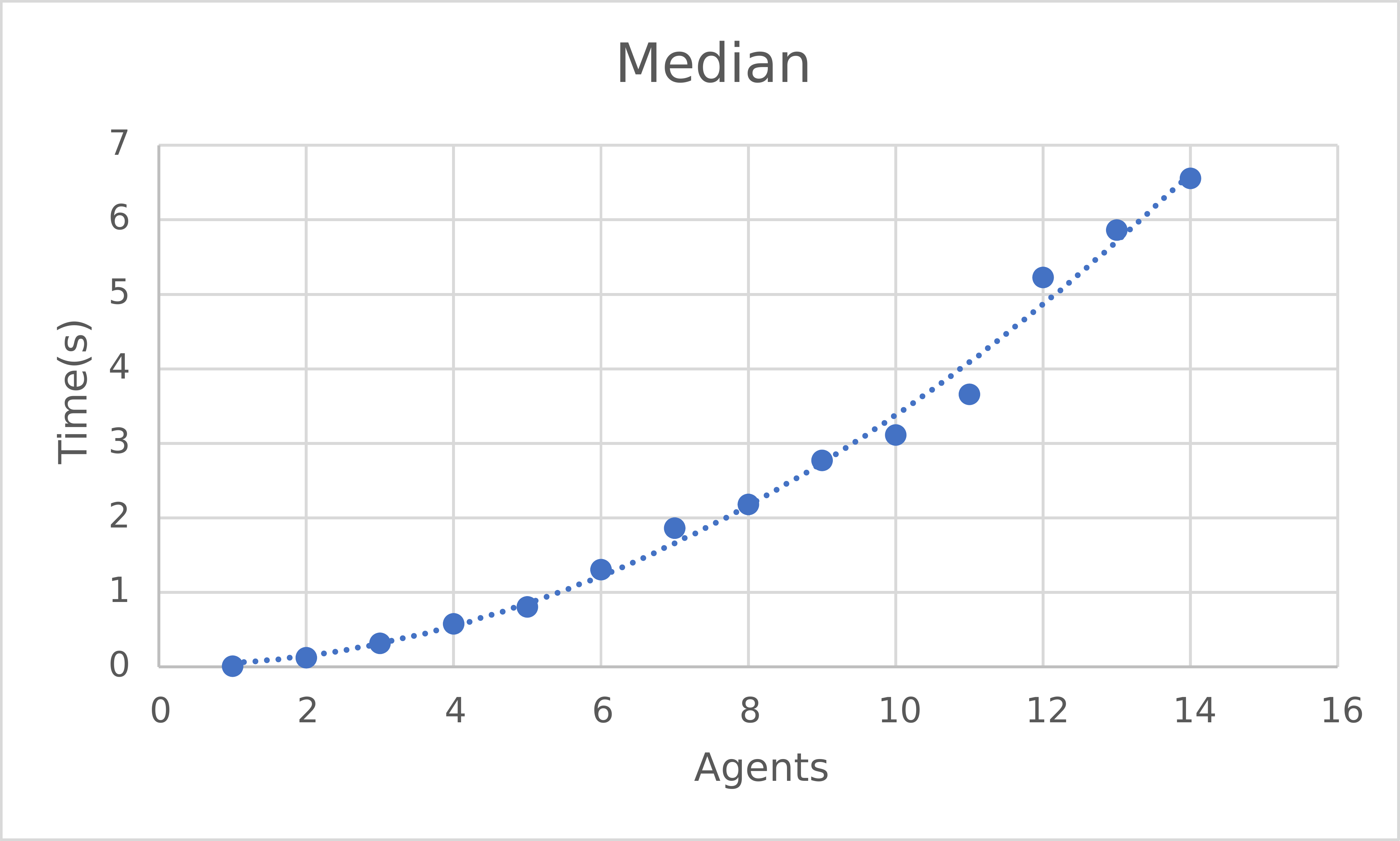}
\centering
\includegraphics[width=0.49\textwidth, height=2.7cm]{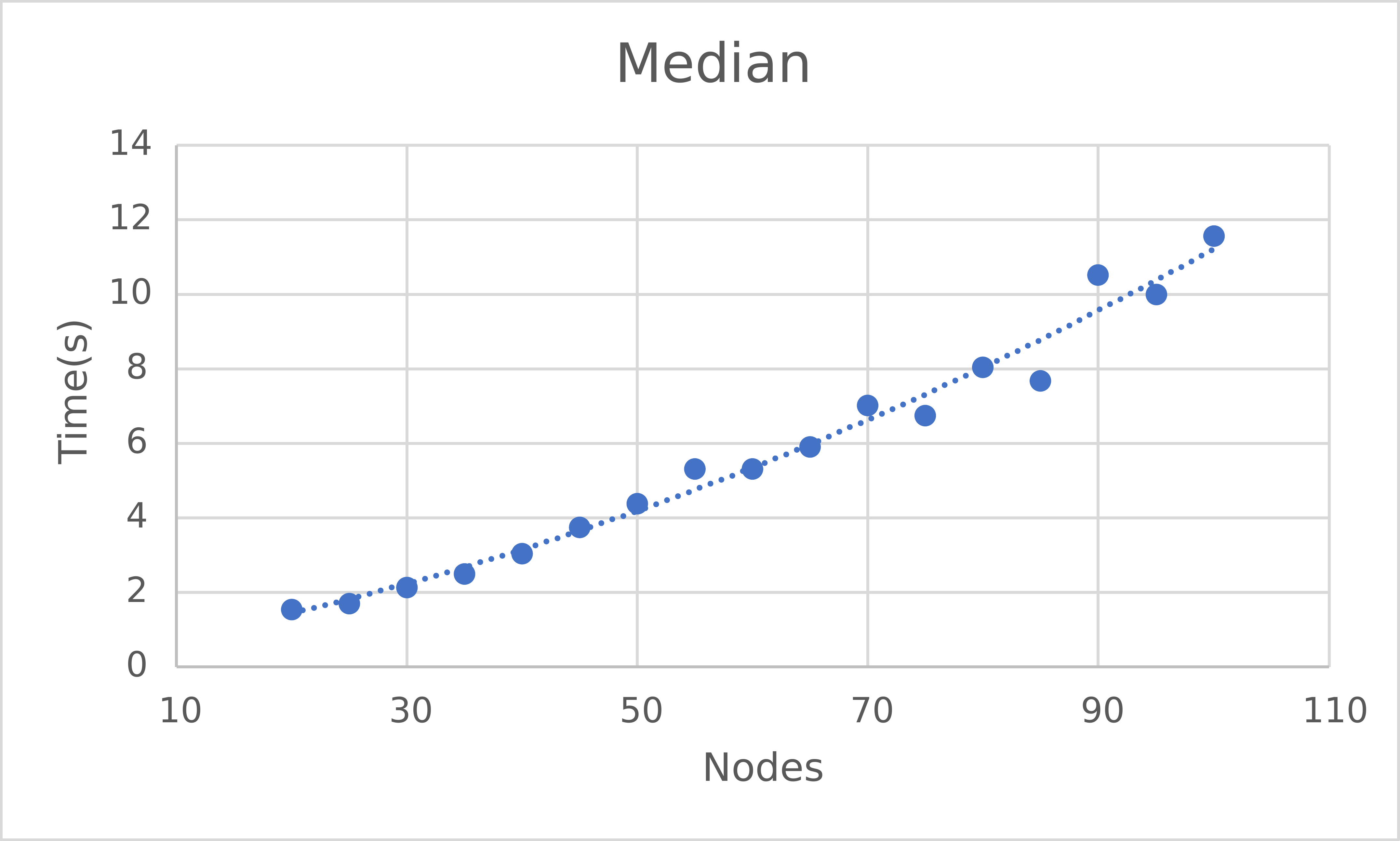}
\end{minipage}
\caption{Median of running times per n. of agents and nodes.}
\label{fig:rt}
\end{figure}

Fig.~\ref{fig:rt} shows the medians of the computational time as a function of the number of nodes and, respectively, the number of agents. Roughly, in both cases, the computational time increases quadratically. In these figures, the trendlines are the least squares approximations with second or third order polynomials.
\begin{figure}[h]
\begin{minipage}[t]{0.5\textwidth}
\centering
\includegraphics[width=0.49\textwidth, height=2.7cm]{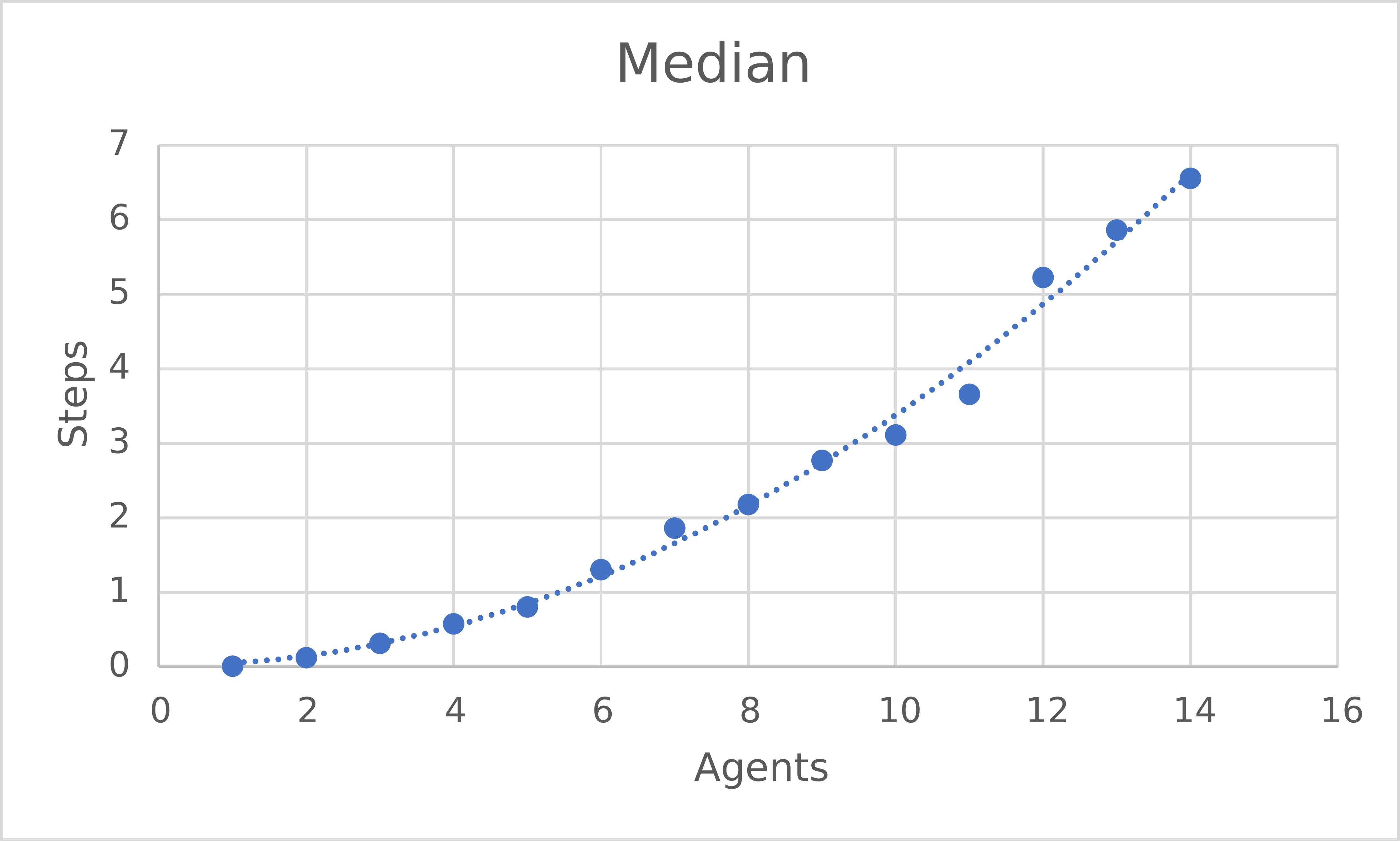}
\centering
\includegraphics[width=0.49\textwidth, height=2.7cm]{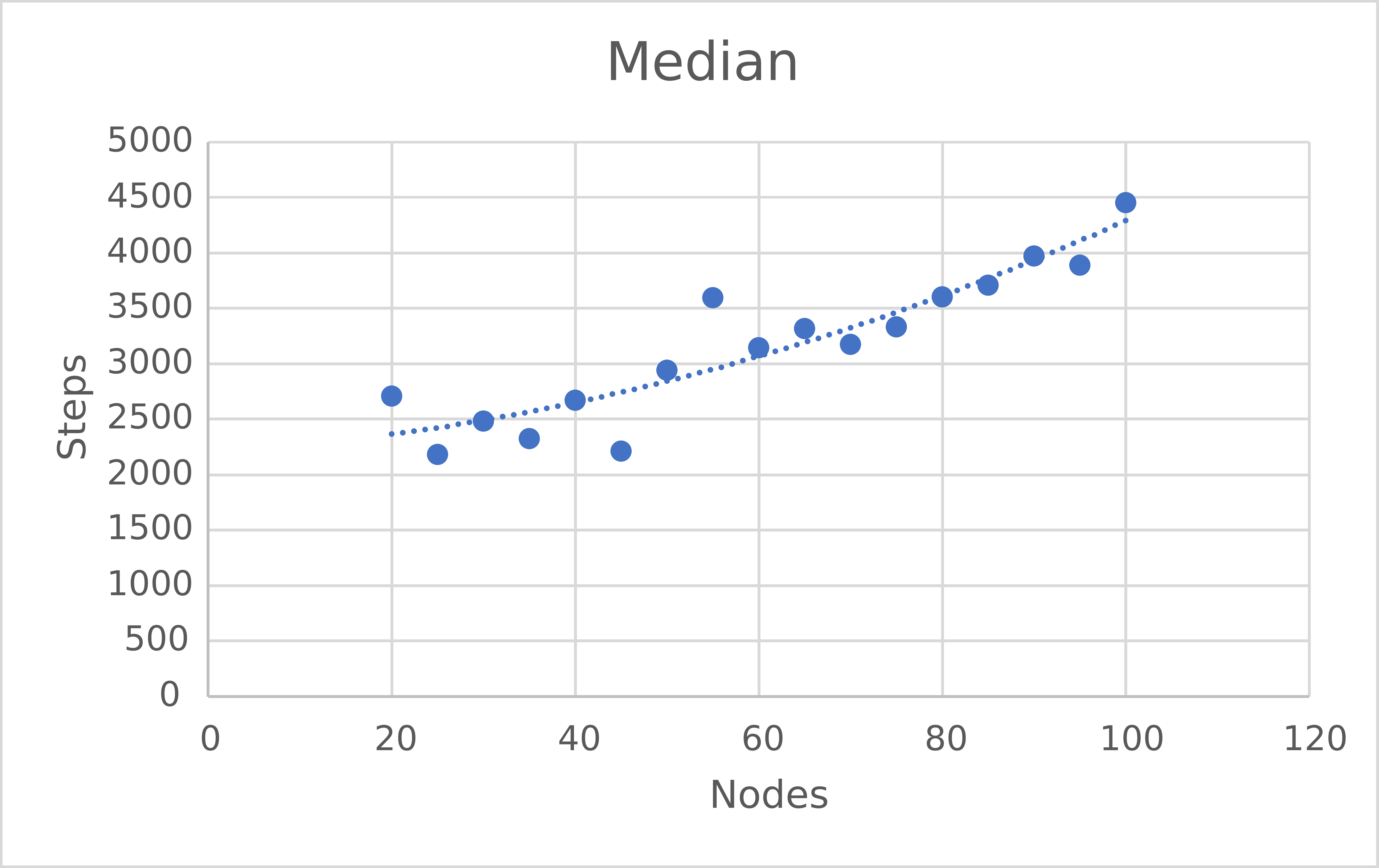}
\end{minipage}
\caption{Median of moves per n. of agents and nodes.}
\label{fig:nm}
\end{figure}                                                                                                                                                                                                                                                                                                                                                                                                             

Fig.~\ref{fig:nm} shows the medians of the overall number of moves as a function of the number of agents and, respectively, the number of nodes. Roughly, the overall number of nodes is a cubic function of the number of agents and a quadratic function of the number of nodes.
\begin{figure}[h]
\begin{minipage}[t]{0.5\textwidth}
\centering
\includegraphics[width=0.5\textwidth]{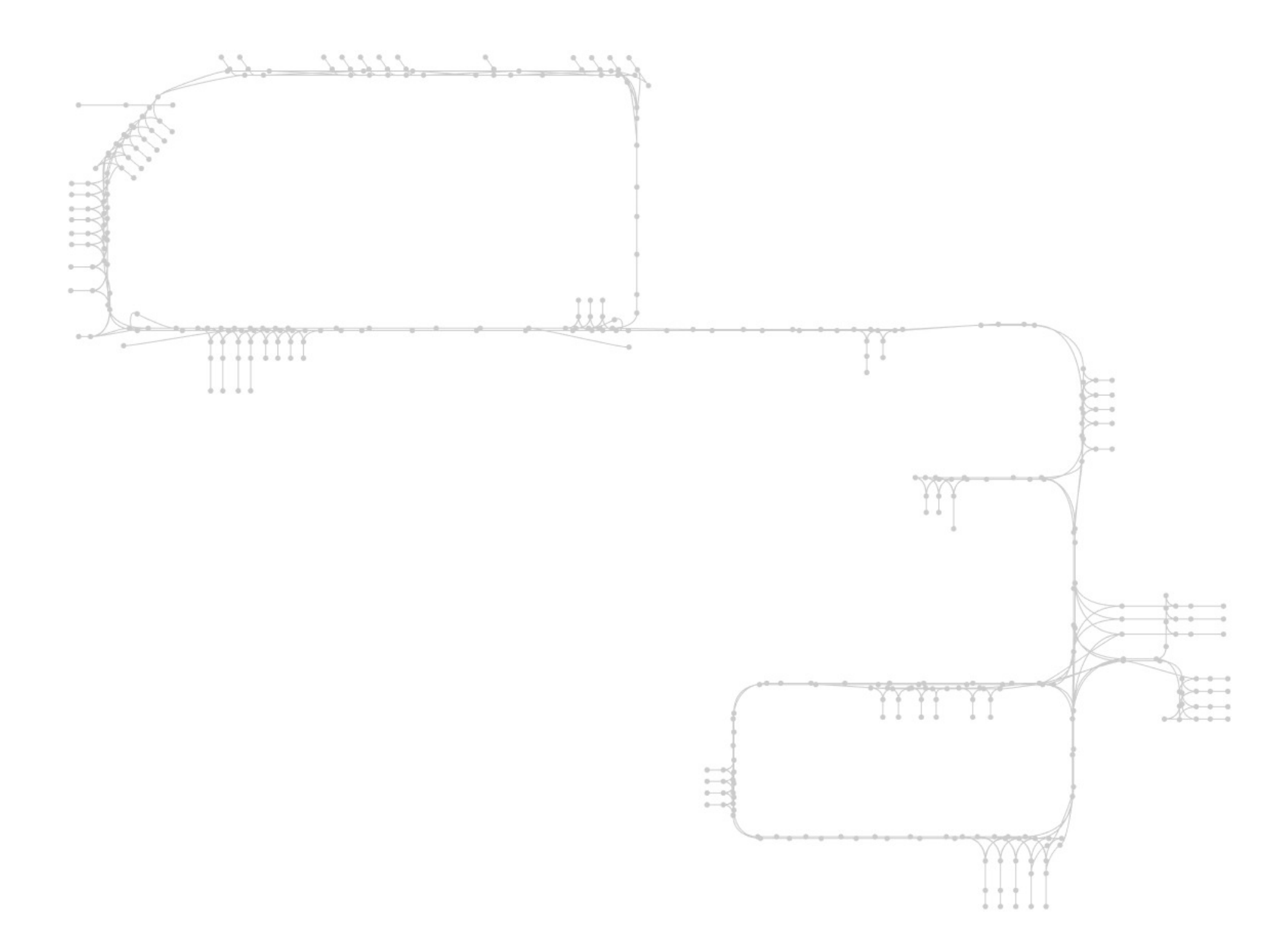}
\end{minipage}
\caption{Graph representing the warehouse.}
\label{fig:pedrollo}
\end{figure}                                                                                                                                                                                                                                                                                                                                                                                                             
We then ran the algorithm for MAPF problem instances on a 397 nodes graph associated to the layout a real warehouse (fig.~\ref{fig:pedrollo}). We ran the algorithm varying the number of agents from 1 to 10.
Also in this case both the number of steps and the running time seem to be increasing in a polynomial way.  
\begin{figure}[h]
\begin{minipage}[t]{0.5\textwidth}
\centering
\includegraphics[width=0.49\textwidth, height=2.7cm]{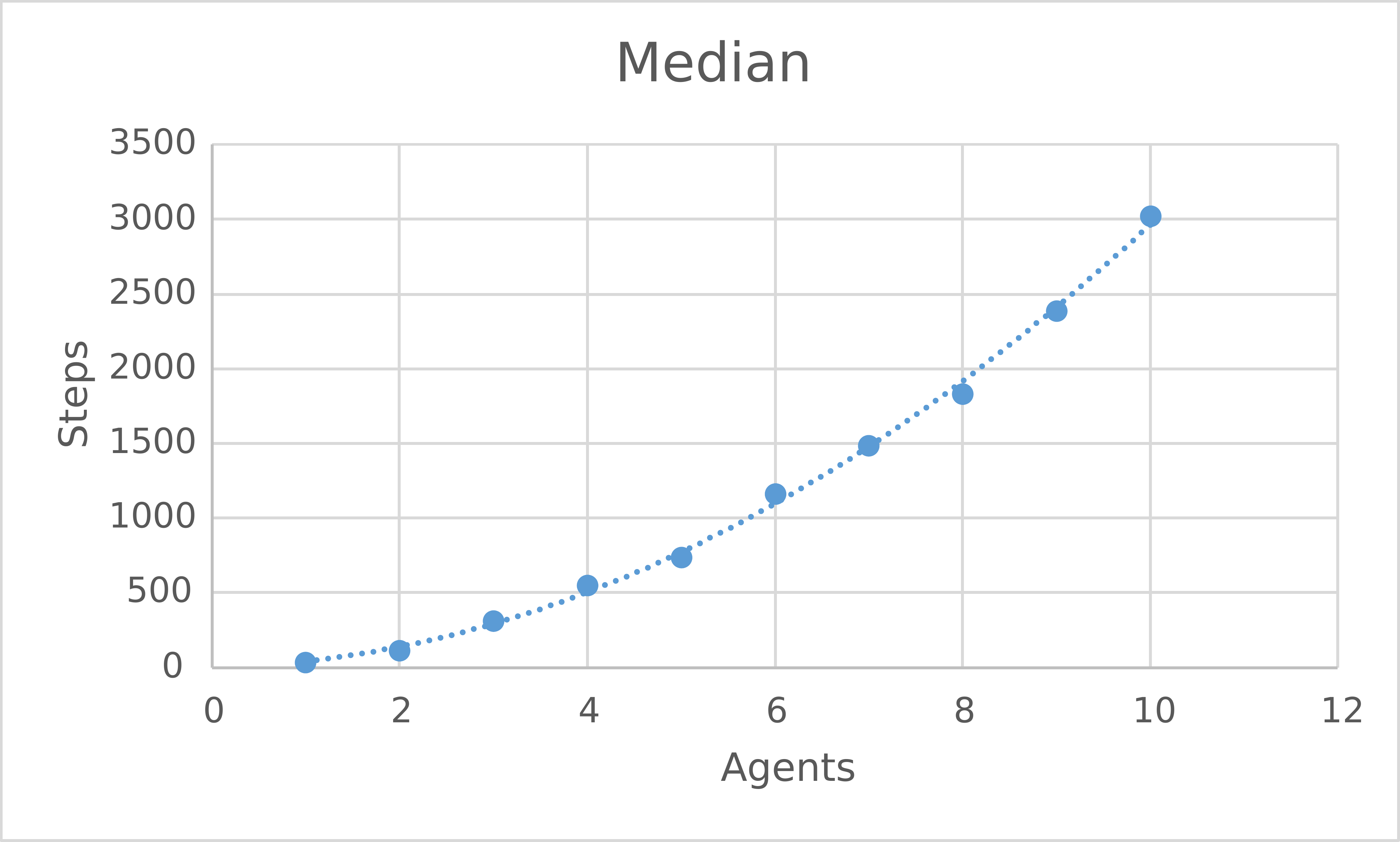}
\centering
\includegraphics[width=0.49\textwidth, height=2.7cm]{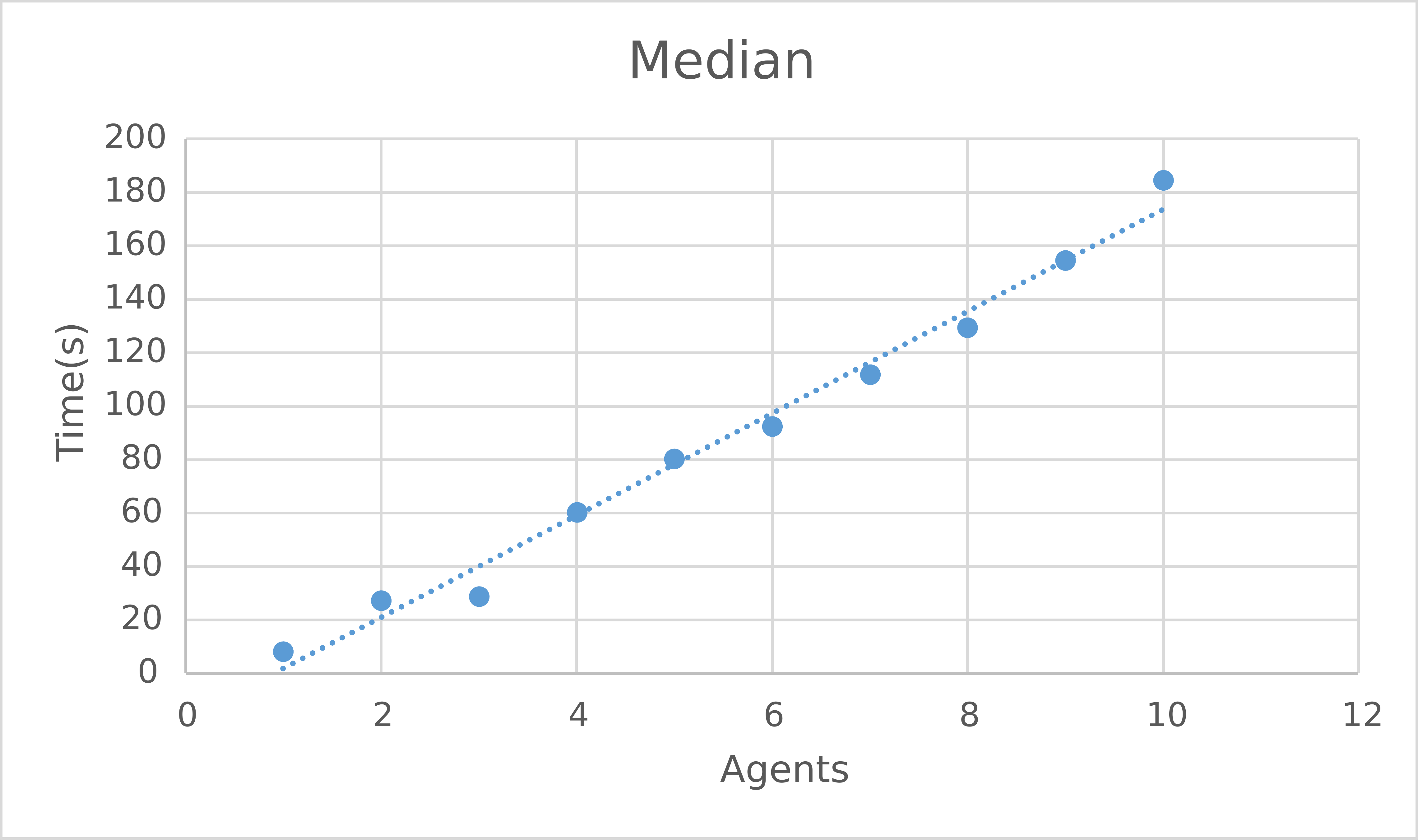}
\end{minipage}
\caption{Median of moves and times per n. of agents on warehouse.}
\label{fig:p_agents}
\end{figure}                                                                                                                                                                                                                                                                                                                                                                                                             
Thus, our simulations confirm the complexity result presented in Theorem~\ref{th:pol}.

\section{Conclusions and Future Work}                                                                                                                                                                     

We proved that the feasibility of MAPF problems on strongly connected digraphs is decidable in linear time (Theorem~\ref{thm_strat_rev}). Moreover, we show that a MAPF problem on a strongly connected digraph is feasible if and only if the corresponding PMT problem on the biconnected component tree is feasible (Corollary~\ref{cor}). Finally, we presented an algorithm (diSC) for solving MAPF problems on strongly connected digraphs in polynomial time with respect to the number of both nodes and agents (Theorem~\ref{th:pol}). As already said, diSC algorithm finds a solution that has often a much larger number of steps than the shortest one. Our next step will be to shorten the solution, for instance by a local search.

\bibliographystyle{siamplain}

\end{document}